\documentclass[11pt]{article}
\usepackage{dsfont}
\usepackage{comment}
\usepackage{amsmath,amssymb,graphicx}
\usepackage{diagbox}
\usepackage{hyperref}
\usepackage[nosort]{cite}
\usepackage{amssymb}
\usepackage{tikz}
\usetikzlibrary{calc} \usetikzlibrary{patterns} \usetikzlibrary{decorations.pathreplacing} \usetikzlibrary{decorations.markings} \usetikzlibrary{decorations.pathmorphing} \usetikzlibrary{positioning}
\usetikzlibrary{matrix}
\usepackage{mathtools}
\usepackage[toc,page]{appendix}
\usepackage{braket}
\usepackage{dirtytalk}

\usepackage{color}
\definecolor{darkred}{rgb}{0.65,0.15,0}
\definecolor{newgreen}{rgb}{0.2,0.62,0.14}
\hypersetup{pdfborder={0 0 0},colorlinks=true,urlcolor=blue,citecolor=blue,linkcolor=darkred,linktocpage=true}

\numberwithin{equation}{section}

\newcommand{\customlabel}[2]{%
	\protected@write \@auxout {}{\string \newlabel {#1}{{#2}{}}}}

\DeclareFontFamily{U}{mathx}{}
\DeclareFontShape{U}{mathx}{m}{n}{ <-> mathx10 }{}
\DeclareSymbolFont{mathx}{U}{mathx}{m}{n}
\DeclareFontSubstitution{U}{mathx}{m}{n}

\DeclareMathAccent{\widecheck}{0}{mathx}{"71}

\def\spa#1.#2{\left\langle#1\,#2\right\rangle}
\def\spb#1.#2{\left[#1\,#2\right]}

\def\zetas{\zeta^*}

\def\beq{\begin{equation}}
\def\eeq{\end{equation}}
\let\Re\relax
\let\Im\relax
\DeclareMathOperator{\Re}{Re}
\DeclareMathOperator{\Im}{Im}

\newcommand{\eq}{\begin{equation}}
\newcommand{\eqe}{\end{equation}}
\newcommand{\eqa}{\begin{eqnarray}}
\newcommand{\eqae}{\end{eqnarray}}

\newcommand{\bea}{\begin{eqnarray}}
\newcommand{\eea}{\end{eqnarray}}
\newcommand{\dd}{\mathrm{d}}

\newcommand{\Es}{E^*}

\newbox\charbox
\newbox\slabox
\def\s#1{{      
        \setbox\charbox=\hbox{$#1$}
        \setbox\slabox=\hbox{$/$}
        \dimen\charbox=\ht\slabox
        \advance\dimen\charbox by -\dp\slabox
        \advance\dimen\charbox by -\ht\charbox
        \advance\dimen\charbox by \dp\charbox
        \divide\dimen\charbox by 2
        \raise-\dimen\charbox\hbox to \wd\charbox{\hss/\hss}
        \llap{$#1$}
}}

\begin{document}

\begin{center}
	\mbox{ }
	\vspace{20mm}

{\bf {\LARGE \sc Integral of depth zero to three basis \\[4mm] of Modular Graph Functions}}

\vspace{6mm}
\normalsize
{\large Mehregan Doroudiani}

\vspace{10mm}

{\it Max-Planck-Institut f\"{u}r Gravitationsphysik (Albert-Einstein-Institut)\\
Am M\"{u}hlenberg 1, 14476 Potsdam, Germany}

\vspace{30mm}

\hrule

\vspace{10mm}

\begin{tabular}{p{14cm}}
Modular Graph Functions (MGFs) are SL(2,$\mathbb{Z}$)-invariant functions that emerge in the study of the low-energy expansion of the one-loop closed string amplitude. To find the string scattering amplitude, we must integrate MGFs over the moduli space of the torus. In this paper, we use the iterated integral representation of MGFs to establish a depth-dependent basis for them, where \say{depth} refers to the number of iterations in the integral. This basis has a suitable Laplace equation. We integrate this basis from depth zero to depth three over the fundamental domain of SL(2,$\mathbb{Z}$) with a cut-off.
\end{tabular}

\vspace{6mm}
\hrule

\end{center}

\thispagestyle{empty}

\newpage
\setcounter{page}{1}

\setcounter{tocdepth}{2}
\tableofcontents
\setcounter{tocdepth}{2}

\newpage 

\section{Introduction}

String scattering amplitudes at one loop are fascinating objects in physics and mathematics. The worldsheet path integral of a one-loop closed string results in an integral over the moduli space of an $n$-punctured torus, where $n$ is the number of external states. This integral can be split into an integral over the configuration of the punctures on a single torus and then an integral over the (bosonic) moduli space of the compact genus-one Riemann surfaces. \textit{Modular graph functions} (MGFs) result from the low-energy expansion of the first integral \cite{Green:1999pv, Green:2008uj}. MGFs are SL($2,\mathbb{Z}$)-invariant functions on the Poincaré upper-half plane, with a lattice-sum representation that can be written by using specific Feynman rules of specific graphs, hence the name. There is a generalization of MGFs, called modular graph forms, which are modular forms rather than functions \cite{DHoker:2016mwo}.

In order to get the low-energy expansion and MGFs, we have to take the inverse of the string tension $\alpha'$ to be small and use the Taylor expansion in the integrand of the first integral. However, this expansion is not valid near the cusp $\tau \rightarrow i\infty$, where $\tau$ is the modulus of the torus. Therefore, we must treat the integral over the configuration space differently near the cusp. As a result, if we want to focus on modular graph forms, we should not integrate them over the regions of moduli space that contain the cusp. This can be done by partitioning the moduli space $\mathcal{M}$ into two segments: $\mathcal{M}_L$, which is away from the cusp, and $\mathcal{M}_R$, which contains the cusp \cite{DHoker:2015gmr}. In the following sections, we will discuss how this splitting is done. The integral over the second part for the four-graviton case in Type II superstring theory was calculated in all orders of $\alpha'$ in \cite{DHoker:2019blr}.

There are different ways to classify MGFs. For example, they can be graded by the transcendental weight $w$. In the study of the four-graviton amplitude in Type II, the integral of MGFs of weight $w$ is related to the $D^{2w}\mathcal{R}^4$ terms in the effective interaction, and all terms up to $D^{12}\mathcal{R}^4$ have already been found \cite{DHoker:2019blr}. Another way to classify the MGFs (or modular graph forms) is by their number of loops \cite{DHoker:2015gmr,Zerbini:2015rss,DHoker:2015wxz}. There are many identities among modular graph forms that can relate MGFs with different numbers of loops \cite{DHoker:2016mwo,DHoker:2016quv,Gerken:2018zcy}. These identities are so helpful that even though five-loop MGFs appear at weight six, knowing the integrals of one-loop and two-loop MGFs is sufficient to integrate MGFs up to weight six \cite{DHoker:2019blr}. One-loop MGFs can be written in terms of (non-holomorphic) Eisenstein series, which have a simple integral using its Poincaré series. The integral of two-loop MGFs was found in \cite{DHoker:2019mib}. See \cite{DHoker:2019blr} for all integrals of MGFs up to weight six. The integral of certain classes of three-loop MGFs was calculated in \cite{DHoker:2021ous}. For a review, refer to \cite{DHoker:2022dxx,Gerken:2020xte}.

There are two main approaches to integrating MGFs. The first one is the unfolding trick, which uses the Poincaré representation of the function. This trick works best when considering the whole fundamental domain $\mathcal{M}$. In \cite{DHoker:2019blr,DHoker:2019mib,DHoker:2021ous}, the Eisenstein series was used to regulate the integral for $\mathcal{M}_L$, which is related to the Rankin-Selberg-Zagier method \cite{rankin_1939, selberg1940bemerkungen,zagier1981rankin}. In this paper, we discuss another view of the problem which does not use the Eisenstein series. This method requires knowing the Poincaré series and the Laurent polynomial of MGFs, which was found in \cite{DHoker:2019txf}. The second approach, which is simpler, involves using the Laplace-Beltrami operator and the Stokes theorem. This method requires Laplace equations with simpler functions which we can integrate as their source. Having a Laplace system for MGFs is very useful but highly non-trivial. For two-loop MGFs, a Laplace system was found in \cite{DHoker:2015gmr}. Additionally, in \cite{Kleinschmidt:2017ege,Basu:2015ayg,Basu:2016xrt,Basu:2019idd}, other Laplace equations were found for different graph topologies. Finding a basis for MGFs with a good Laplace system is challenging, and this technique can be applied in a few situations. These integration methods were also considered in \cite{Angelantonj:2011br,Angelantonj:2012gw,Angelantonj:2013eja,Pioline:2014bra} for different cases of string one-loop amplitudes.

This paper uses a combination of both approaches to find the integral over $\mathcal{M}_L$. Our guide to finding the basis for MGFs with a suitable Laplace system is their generating function \cite{Gerken:2019cxz,Gerken:2020yii}. The generating function provides an iterated integral representation of the MGFs. For the iterated integrals, one can use depth filtration instead of weight grading, where depth refers to the number of iterations. The generating function can be expressed as a generating series of functions called $\beta^{\rm sv}$ (or their modular form counterparts $\beta^{\rm eqv}$) with a matrix representation of the Tsunogai derivation algebra \cite{Gerken:2020yii,Mafra:2019ddf, Mafra:2019xms, Dorigoni:2022npe}. Any modular graph form can be expanded as a certain linear combination of  $\beta^{\rm sv}$'s. Since these functions have known Cauchy-Riemann equations, it is possible to write a basis for them with a suitable Laplace system. This was achieved for depth two at arbitrary weight in \cite{Dorigoni:2021jfr,Dorigoni:2021ngn}, and for a variety of depth three examples in \cite{drewitt2022laplace,Dorigoni2023:appear}. It is worth mentioning that the space spanned by this basis is larger than the space of modular graph forms and also includes cusp forms. Tsunogai's derivation algebra projects the space spanned by this basis to modular graph forms \cite{Dorigoni:2022npe,Dorigoni:2021jfr,Dorigoni:2021ngn,drewitt2022laplace,Dorigoni2023:appear}.

\section{Review on modular forms and fundamental domain}

We consider a torus with modulus $\tau$, which indicates the shape of the torus. $\tau$ is a parameter on the Poincar\' e upper half plane $\mathcal{H}$ and we write it as $\tau = \tau_1 + i \tau_2$ where both $\tau_1$ and $\tau_2$ are real numbers and $\tau_2 > 0$. The mapping class group of a torus is the modular group ${\rm SL}(2,\mathbb{Z})$.

\begin{align}
	\label{SL2Z}
	\operatorname{SL}(2, \mathbb{Z})=\left\{\gamma = \left(\begin{array}{ll}
		a & b \\
		c & d
	\end{array}\right): a, b, c, d \in \mathbb{Z}, a d-b c=1\right\} \, ,
\end{align}

\noindent and under this group, the modulus $\tau$ transforms as below

\begin{align}
	\label{modTau}
	\tau \rightarrow \gamma \cdot \tau = \frac{a\tau+b}{c\tau+d} \, ,
\end{align}

\noindent  and we use the closure of the fundamental domain ${\rm PSL}(2,\mathbb{Z}) \backslash \mathcal{H}$, as the moduli space of the torus

\begin{align}
	\label{FunDom}
	\mathcal{M}=\left\{\tau \in \mathcal{H}:|\operatorname{Re}(\tau)| \leq \frac{1}{2},|\tau| \geq 1\right\} \, .
\end{align}

\begin{figure}[h!]
	\centering
	\begin{tikzpicture}

		\def\L{2} 
		
		\fill[gray!20] (1/2,0.866025) -- (1/2,\L) -- (-1/2,\L)--(-1/2,0.866025) -- (1,0.866025);
		\fill[white] (1,0) arc (0:180:1) -- cycle;
		
		\draw (1,0) arc (0:180:1);
		
		\draw (1/2,0.866025) -- (1/2,\L) -- (-1/2,\L)--(-1/2,0.866025);
		\draw (1/2,0.866025) -- (1/2,3);
		\draw (-1/2,0.866025) -- (-1/2,3);
		\node at (1,\L+0) [left] {$L$};
		\node at (-1,\L-0.6) [left] {$\mathcal{M}_L$};
		\node at (-1,\L+0.6) [left] {$\mathcal{M}_R$};
		
		 \draw[->]   (-0.1,\L-0.6) --  (-1,\L-0.6);
		  \draw[->]   (-0.1,\L+0.6) --  (-1,\L+0.6);
		\draw[->] (-2,0) -- (2,0) node[right] {$\tau_1$};
		\draw[->] (0,0) -- (0,\L+1) node[above] {$\tau_2$};
		
	\end{tikzpicture}
	\caption{Fundamental domain with a cut-off at $\tau_2 = L$. The shaded region is $\mathcal{M}_L$ and the complement is $\mathcal{M}_R$}
	\label{ML}
\end{figure}
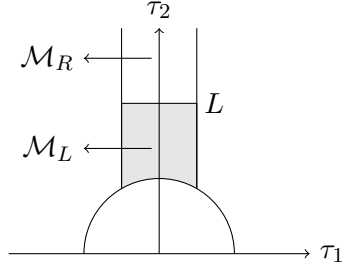

\noindent We split the fundamental domain into two parts $\mathcal{M} = \mathcal{M}_L \cup \mathcal{M}_R$ where $\mathcal{M}_R=\mathcal{M} \cap\{\operatorname{Im}(\tau)>L\}$ and $\mathcal{M}_L=\mathcal{M} \cap\{\operatorname{Im}(\tau)\leq L\}$ for some cut-off $L>1$. Figure \ref{ML} shows the schematic of $\mathcal{M}_L$ which is the region we want to integrate over.

\noindent For the integration over this region, we use the modular-invariant Poincar\' e metric $\dd \mu$, where

\begin{align}
	\label{dmu}
	\dd \mu = \frac{\dd^2 \tau}{\tau_2^2} = \frac{i \dd\tau \wedge \dd\bar\tau}{2\tau_2^2} \, ,
\end{align}

\noindent and by using this metric, the volume of the fundamental domain is

\begin{align}
	\label{VolumeDom}
	\int_{\mathcal{M}} \dd\mu = \frac{\pi}{3} \, .
\end{align}

\noindent A modular form $F(\tau)$ is a complex-valued function on the upper-half plane with a specific transformation under the ${\rm SL}(2,\mathbb{Z})$ action (\ref{SL2Z})

\begin{align}
	\label{modTransform}
	F(\gamma \cdot \tau) = (c\tau +d)^w (c\bar\tau+d)^{w^*} F(\tau) \, ,
\end{align}

\noindent where $w, w^* \in \mathbb{R}$ and $w-w^*\in \mathbb{Z}$. In this paper, $F(\tau)$ is not necessarily a holomorphic function. $w$ and $w^*$ are called the holomorphic and the anti-holomorphic modular weights of $F(\tau)$ and we denote it as weight ($w,w^*$). A modular form with $w=w^*=0$ is called a modular function.

A very important modular function in this paper is the non-holomorphic Eisenstein series. It has a lattice-sum representation as follows

\begin{align}
	\label{EisSum}
	E_s(\tau)=\left(\frac{\tau_2}{\pi}\right)^s\sum_{(m,n) \in \mathbb{Z}^2\backslash (0,0)} \frac{1}{|m\tau+n|^{2 s}} \, ,
\end{align}

\noindent and also the following Fourier expansion

\begin{equation}
	\label{EisFourier}
	E_s(\tau)=c_s\tau_2^s+\tilde{c}_s\tau_2^{1-s}+\frac{4\sqrt{\tau_2}}{\Gamma(s)}\sum_{N\neq0}|N|^{s-1/2}\sigma_{1-2s}(N)K_{s-1/2}(2\pi |N|\tau_2)e^{2\pi iN\tau_1} \, .
\end{equation}

\noindent $K_{\alpha}(x)$ is the modified Bessel function of the second kind and

\begin{align}
	\label{EisConst}
	\begin{split}
	&c_s  = \frac{2\zeta(2s)}{\pi^s} \, ,
	\\
	&\tilde{c}_s = \frac{2\Gamma(s-\frac{1}{2})\zeta(2s-1)}{\Gamma(s)\pi^{s-1/2}} \, ,
	\end{split}
\end{align}

\noindent where $\zeta(s)$ is the Riemann zeta function. The other convention that is more useful for us is the completed non-holomorphic Eisenstein series

\begin{equation}
	\label{EtoEstar}
 E^*_s = \frac{\Gamma(s)}{2} E_s \, ,
 \end{equation}
 which satisfies the functional relation $E^*_{s}(\tau) = E^*_{1-s}(\tau)$ with simple poles at $s=0$ and $s=1$ and its Fourier expansion is

\begin{align}
	\label{compEis}
	E^*_s(\tau)= & \, \zeta^*(2s)\tau_2^s+\zeta^*(2s-1)\tau_2^{1-s} \notag \\
	&+2\sqrt{\tau_2}\sum_{N\neq0}|N|^{s-1/2}\sigma_{1-2s}(N)K_{s-1/2}(2\pi |N|\tau_2)e^{2\pi iN\tau_1} \, ,
\end{align}

\noindent where $\zeta^*$ is the completed zeta function and is defined as follows

\begin{equation}
	\label{zetastar}
	\zeta^*(s) = \pi^{-s/2}\Gamma(s/2)\zeta(s) \, .
\end{equation}

\noindent $\zetas$ also has a functional relation $\zeta^*(s) = \zeta^*(1-s)$ and simple poles at $s=1$ and $s=0$.

For convenience, we may also use the shortened notation

\begin{equation}
	\label{compEisShort}
	E^*_s(\tau) = \zetas(2s)\tau_2^s+\zetas(2s-1)\tau_2^{1-s}+\mathcal{E}_s(\tau)=e_s(\tau_2)+\mathcal{E}_s(\tau) \, .
\end{equation}

\noindent $\mathcal{E}_s(\tau)$ is an entire function of $s$ and it is a rapid decay function with respect to $\tau_2$. In this context, rapid decay means it vanishes at the cusp faster than any polynomial. Furthermore, $\zetas(2)=\frac{\pi}{6}$ and $\underset{s=1}{\text{Res}}(\Es_s(\tau))=\frac{1}{2}$. Fourier expansion of $\mathcal{E}_s(\tau)$ is written as

\begin{equation}
	\label{EisNonZero}
	\mathcal{E}_s(\tau) = \sum_{m\neq0}\mathcal{E}_{s,m}(\tau_2)e^{2m\pi i \tau_1} \, ,
\end{equation}

\noindent which introduces the notation $\mathcal{E}_{s,m}(\tau_2)$ for non-holomorphic Eisenstein's Fourier coefficients w.r.t $\tau_1$. 

Now we consider a general modular function $F(\tau)$ with the following Fourier expansion

\begin{equation}
	\label{GenFourier}
	F(\tau) = a_0(\tau_2) + \sum_{m\neq0} A_m(\tau) = a_0(\tau_2) + \sum_{m\neq0} a_m(\tau_2)e^{2m\pi i \tau_1} \, .
\end{equation}

\noindent Assume that the zero mode has a Laurent polynomial and an exponentially suppressed part, so for $a_0(\tau_2)$ we have

\begin{equation}
	\label{GenZeroMode}
	a_0(\tau_2) = \sum_{n\in I} \alpha_n \tau_2^n + {R(\tau_2)} \, ,
\end{equation}

\noindent  where $I$ is a finite set of real numbers. In this paper, $R(\tau_2)$ is called the rapid decay part, the non-perturbative or exponentially suppressed term of the zero mode. The Mellin transform of the rapid decay part is important for us and equals to

\begin{equation} 
	\label{MelTr}
	M(F(\tau);s) = \int_{0}^{\infty} R(\tau_2)\zetas(2s)\tau_2^{s-2} \dd\tau_2  \, ,
\end{equation}

\noindent which is also called the regularized Rankin-Selberg transform \cite{rankin_1939, selberg1940bemerkungen}. This integral is convergent for large enough ${\rm Re}(s)$ and by meromorphic continuation, we define

\begin{equation}
	\label{MelRes}
	M(F(\tau)) = \underset{s=1}{\text{Res}}(M(F(\tau);s)) \, .
\end{equation}

In the following sections, we go through different integration methods to a modular function $F(\tau)$ over $\mathcal{M}_L$, and finally, we start to integrate the basis of MGFs.

\section{Unfolding trick on $\mathcal{M}_L$}

The first method that we study is the unfolding. The idea behind the unfolding is to integrate over a much simpler region, a rectangle. The unfolding trick was made for functions with rapid decay zero modes that guarantee the integral's convergence over the whole fundamental domain $\mathcal{M}$. In \cite{DHoker:2019mib,DHoker:2021ous}, this method was used by taking out the Laurent polynomial using the non-holomorphic Eisenstein series, which is a description in Zagier's regularization of Rankin-Selberg method \cite{zagier1981rankin}.

In this section, we study how to use the unfolding trick directly for $\mathcal{M}_L$. In order to do that, we consider the Borel subgroup $\Gamma_{\infty}$ of ${\rm SL}(2,\mathbb{Z})$ 
 
 \begin{align}
 	\label{GammaInf}
 	\Gamma_{\infty}=\left\{ \pm\left(\begin{array}{cc}
 		1 & n \\
 		0 & 1
 	\end{array}\right): n \in \mathbb{Z}\right\} \subset \mathrm{SL}(2, \mathbb{Z}) \, .
 \end{align}

\noindent The action of this group on $\tau$ translates it by $n$, therefore it stabilizes the cusp $\tau=i\infty$. Using the Borel subgroup, Poincar\' e series of a modular function $F(\tau)$ can be written as

\begin{equation}
	\label{PoincareSum}
	F(\tau) = \sum_{\gamma \in \Gamma_{\infty}\backslash {\rm SL}(2,\mathbb{Z})}f(\gamma\cdot \tau) \, ,
\end{equation}

\noindent where $f(\tau)$ is called the seed function. $f(\tau)$ is invariant under $\tau \rightarrow \tau+1$. What the unfolding trick does is to change the integration domain from ${\rm PSL}(2,\mathbb{Z}) \backslash \mathcal{H}$ to $\Gamma_{\infty} \backslash \mathcal{H}$ and since

\begin{align}
	\label{HoverGammaInf}
	\Gamma_{\infty} \backslash \mathcal{H}=\{\tau \in \mathcal{H}: 0 \leq \operatorname{Re}(\tau) \leq 1\} \, ,
\end{align}

\noindent the integration domain becomes simpler since this region is just a strip, at the cost of replacing $F(\tau)$ by $f(\tau)$. Then we have the following theorem which can be seen as a variant of the Rankin-Selberg-Zagier integral \cite{rankin_1939,selberg1940bemerkungen,zagier1981rankin}.

\textbf{\hypertarget{Th1}{Theorem 1}:} The integral of a modular function $F(\tau)$ using its seed function $f(\tau)$ is

\begin{equation}
	\label{ThOne}
	\int_{\mathcal{M}_L}F(\tau)\dd\mu = \int_0^L f_0(\tau_2)\frac{\dd\tau_2}{\tau_2^2}-\int_{L}^{\infty}(a_0(\tau_2)-f_0(\tau_2))\frac{\dd\tau_2}{\tau_2^2} \, ,
\end{equation}

\noindent where $f_0(\tau_2)$ is the zero mode of the seed function.

\textbf{Proof:}  For integrating over $\mathcal{M}_L$ we need to define another region $\mathcal{H}_L$ where

\begin{equation}
	\mathcal{H}_L = \{\tau\in\mathcal{H} | \text{Im}(\tau)\leq L\} - \bigcup_{c\geq1}\bigcup_{\substack{a\in\mathbb{Z} \\ (a,c)=1}} S_{a/c} \, ,
\end{equation}

\noindent which satisfies the following relation

\begin{align}
	\label{HL}
	\mathcal{M}_L = {\rm PSL}(2,\mathbb{Z}) \backslash \mathcal{H}_L\, ,
\end{align}

\noindent where $S_{a/c}$ is a disc of radius $\frac{1}{2c^2L}$ tangent to the real axis at $\frac{a}{c}$ \cite{zagier1981rankin}. By replacing the Poincar\'e sum expansion (\ref{PoincareSum}) as $F(\tau)$ in the integral, we have

\begin{align}
	\begin{split}
		\int_{\mathcal{M}_L} F(\tau)\dd\mu &= \int_{\Gamma_{\infty}\backslash\mathcal{H}_L} f(\tau)\dd\mu\\
		&= \int_0^L\frac{\dd\tau_2}{\tau_2^2}\int_0^1 \dd\tau_1 f(\tau) - \sum_{c=1}^{\infty}\sum_{\substack{a(\text{mod}\, c) \\ (a,c)=1}}\int_{S_{a/c}}f(\tau)\dd\mu \, ,
	\end{split}
\end{align}

\noindent and we transform the last integral by acting the element $\gamma_0=\begin{pmatrix}
	a & b\\
	c & d
\end{pmatrix}\in {\rm PSL}(2,\mathbb{Z})$ and the fact that $\gamma_0\cdot S_{a/c} = \{\tau\in\mathcal{H}|\tau_2\geq L\}$.

\begin{align}
	\int_{S_{a/c}}f(\tau)\dd\mu &= \int_L^{\infty}\frac{\dd\tau_2}{\tau_2^2}\int_{-\infty}^{\infty}\dd\tau_1 f(\gamma_0\cdot \tau) \notag \\
	&= \int_L^{\infty}\frac{\dd\tau_2}{\tau_2^2}\int_0^1\dd\tau_1\sum_{n=-\infty}^{\infty}f(\gamma_0\cdot (\tau+n)) \notag \\
	&= \int_{\mathcal{M}-\mathcal{M}_L}\sum_{\gamma=\big(\begin{smallmatrix}
			a & \bullet\\
			c & \bullet
		\end{smallmatrix}\big)} f(\gamma\cdot \tau)\dd\mu \, .
\end{align}

\noindent Finally, summing over all $c>0$ and $a$ (mod $c$) gives us the full Poincar\' e representation of $F(\tau)$ except the $c=0$ term

\begin{equation}
	\int_{\mathcal{M}_L}F(\tau)\dd\mu = \int_0^L f_0(\tau_2)\frac{\dd\tau_2}{\tau_2^2}-\int_{L}^{\infty}(a_0(\tau_2)-f_0(\tau_2))\frac{\dd\tau_2}{\tau_2^2} \, ,
\end{equation}

\noindent where $f_0(\tau_2)$ is the Fourier zero mode of the seed

\begin{equation}
	f_0(\tau_2) = \int_0^1 f(\tau)\dd\tau_1 \, ,
\end{equation}

\noindent and the proof is done.

\section{Rankin-Selberg-Zagier method on $\mathcal{M}_L$}

The Rankin-Selberg method \cite{rankin_1939,selberg1940bemerkungen} uses the unfolding method on the function $E^*_s(\tau) F(\tau)$ for some $s$ and then takes the limit $s\rightarrow 1$ where $E^*_s(\tau)$ has a simple pole. Then, the residue of the result gives us the integral of $F(\tau)$. But the Rankin-Selberg method only works when the zero mode of $F(\tau)$ is a rapid decay function of $\tau_2$. In \cite{zagier1981rankin}, Zagier proposed a regularization scheme for functions with Laurent polynomial and logarithmic terms in their zero mode. In this method, $\mathcal{M}$ is partitioned into two segments $\mathcal{M}_L$ and $\mathcal{M}_R$ precisely as we did here, then the integral over $\mathcal{M}_R$ is treated differently in a way that resumming the integral of the two parts is a finite number which is independent of $L$. Since the integral of MGFs over $\mathcal{M}$ is also not convergent, Zagier's regularization has many applications in this topic. Nevertheless, here, since we already know that the integral of the amplitude must be treated differently over $\mathcal{M}_R$ and the MGF expansion does not hold there, we do not try to take MGF integrals over the whole fundamental domain, and we are interested to know how to integrate them over $\mathcal{M}_L$, therefore we do not use any regularization. The following theorems are similar to the Rankin-Selberg-Zagier methods \cite{rankin_1939,selberg1940bemerkungen,zagier1981rankin}, but only consider the integral over $\mathcal{M}_L$. 

\textbf{\hypertarget{Th2}{Theorem 2}:} The integral of a modular function $F(\tau)$ over $\mathcal{M}_L$ is

\begin{align}
	\label{ThTwo}
	\int_{\mathcal{M}_L}  F(\tau)\dd\mu =& \frac{\pi}{3} \alpha_0 + \alpha_1 \big(-\gamma_E + \log(4\pi L)\big)\\
	&+2 M(F)+\sum_{n\in I\backslash \{1\}} \alpha_n \frac{L^{n-1}}{n-1} {-\int_L^{\infty}R(\tau_2)\frac{\dd\tau_2}{\tau_2^2}} \, , \notag
\end{align}

\noindent where $M(F(\tau))$ is the residue of the regularized Rankin-Selberg transform (\ref{MelRes}), $\alpha_n$'s are the coefficients of the Laurent polynomial and $R(\tau)$ is the rapid decay term of the zero mode (\ref{GenZeroMode}).  Also, $\gamma_E$ is the Euler-Mascheroni constant.

\noindent Note that the last term is a rapid decay function of $L$.

\textbf{\hypertarget{Th3}{Theorem 3}:} For large ${\Re}(s)$ (large is defined in the proof) the integral of a modular function $F(\tau)$ multiplied by a non-holomorphic Eisenstein series over $\mathcal{M}_L$ is

\begin{align}
	\label{ThThree}
	\begin{split}
		\int_{\mathcal{M}_L}F(\tau)\Es_s(\tau)\dd\mu =& M(F(\tau);s) + \zetas(2s) h_L(s)+\zetas(2s-1) h_L(1-s)\\
		&- \int_L^{\infty}R(\tau_2)e_s(\tau_2)\frac{\dd\tau_2}{\tau_2^2}+ \int_L^{\infty}\bigg(\sum_{m\neq 0} \mathcal{E}_{s,m}(\tau_2)a_{-m}(\tau_2)\bigg)\frac{\dd\tau_2}{\tau_2^2} \, ,
	\end{split}
\end{align}

\noindent where

\begin{equation}
	\label{hL}
	h_L(s)=\sum_{n\in I} \alpha_n \frac{L^{n+s-1}}{n+s-1} \, .
\end{equation}

\noindent We used the expansions (\ref{compEis}) to (\ref{GenZeroMode}). Theorem \hyperlink{Th2}{2} is a corollary of theorem \hyperlink{Th3}{3}. 

\textbf{Proof of theorem \hyperlink{Th2}{2} and  \hyperlink{Th3}{3}: }We start by writing the Poincar\' e series of $E^*_s(\tau)$

\begin{align}
	\label{EisUnfold}
	E^*_s(\tau) = \zetas(2s) \sum_{\gamma \in \Gamma_{\infty} \backslash {\rm PSL}(2,\mathbb{Z})} \Im (\gamma\cdot\tau)^s \, ,
\end{align}

\noindent and then we unfold $E^*_s(\tau)$ in $E^*_s(\tau) F(\tau)$

\begin{align}
	\begin{split}
		\int_{\mathcal{M}_L} F(\tau)\Es_s(\tau)\dd\mu &= \int_{\Gamma_{\infty}\backslash\mathcal{H}_L} F(\tau)\zetas(2s)\tau_2^s \dd\mu\\
		&= \int_0^L\frac{\dd\tau_2}{\tau_2^2} \int_0^1 \dd\tau_1 F(\tau)\zetas(2s)\tau_2^s - \sum_{c=1}^{\infty} \, \sum_{\substack{a(\text{mod}\, c) \\(a,c)=1}} \int_{S_{a/c}}F(\tau){\zetas(2s)}\tau_2^s \dd\mu \, .
		\end{split}
\end{align}

\noindent By acting $\gamma_0=\begin{pmatrix}
	a & b\\
	c & d
\end{pmatrix}\in \Gamma$ on the last integral for all $b$ and $d$ allowed, we have $\gamma_0\cdot S_{a/c} = \{\tau\in\mathcal{H}|\tau_2\geq L\}$ and

\begin{align}
	\begin{split}
	\int_{S_{a/c}}F(\tau)\tau_2^s \dd\mu &= \int_L^{\infty}\frac{\dd\tau_2}{\tau_2^2}\int_{-\infty}^{\infty}\dd\tau_1 F(\tau)\text{Im}(\gamma_0\cdot\tau)^s \\
	&= \int_L^{\infty}\frac{\dd\tau_2}{\tau_2^2}\int_0^1 \dd\tau_1 F(\tau)\sum_{n=-\infty}^{\infty}\text{Im}\left(\gamma_0\cdot(\tau+n)\right)^s \\
	&= \int_{\mathcal{M}-\mathcal{M}_L}F(\tau)\sum_{\gamma=\big(\begin{smallmatrix}
			a & \bullet\\
			c & \bullet
		\end{smallmatrix}\big)} \text{Im}(\gamma\cdot\tau)^s \dd\mu \, .
	\end{split}
\end{align}

\noindent Summing over all $c>0$ and $a$ (mod $c$) gives us the full Poincar\' e representation of Eisenstein series except the $c=0$ term

\begin{equation}
	\int_{\mathcal{M}_L}F(\tau)\Es_s(\tau) \dd\mu = \zetas(2s)\int_0^La_0(\tau_2)\tau_2^{s-2}\dd\tau_2-\int_{\mathcal{M}-\mathcal{M}_L}F(\tau)\left(\Es_s(\tau)-\zetas(2s)\tau_2^s\right) \dd\mu \, .
\end{equation}

{
	\noindent We can write the first term as
	\begin{align}
		\begin{split}
		\int_0^La_0(\tau_2)\tau_2^{s-2}\dd\tau_2 &= \int_0^L R(\tau_2)\tau_2^{s-2}\dd\tau_2+\sum_{n \in I} \alpha_n \int_0^L\tau_2^{n+s-2}\dd\tau_2\\
		&= \int_0^{\infty}R(\tau_2)\tau_2^{s-2}\dd\tau_2-\int_L^{\infty}R(\tau_2)\tau_2^{s-2}\dd\tau_2+h_L(s) \, ,
		\end{split}
	\end{align}
	
	\noindent where we got ${\rm Re}(s)$ large enough to make sure that the last term is not troublesome ($h_L(s)$ is given in (\ref{hL})). In other words, ${\rm Re}(s)>3-\min(I)$. The first term on the right hand side is $M(F(\tau);s)/\zetas(2s)$.
}

By using the Fourier expansion of the non-holomorphic Eisenstein series, we have

\begin{align}
	\begin{split}
		\int_{\mathcal{M}_L}F(\tau)\Es_s(\tau)\dd\mu =& \int_0^L a_0(\tau_2)\zetas(2s)\tau_2^{s-2}\dd\tau_2-\int_L^{\infty}a_0(\tau_2)\zetas(2s-1)\tau_2^{-1-s}\dd\tau_2\\
		&-\int_L^{\infty}\bigg(\int_0^1F(\tau)\mathcal{E}_s(\tau)d\tau_1\bigg)\frac{\dd\tau_2}{\tau_2^2} \, .
	\end{split}
\end{align}

\noindent The integrals are convergent for large enough ${\Re}(s)$ and we can continue their results analytically. Similarly we have
{
	
	\begin{equation}
		\int_L^{\infty} a_0(\tau_2)\tau_2^{-s-1}\dd\tau_2 = \int_L^{\infty}R(\tau_2)\tau_2^{-s-1}\dd\tau_2-h_L(1-s) \, .
	\end{equation}
	\noindent Writing this all together
	\begin{align}
		\begin{split}
			\int_{\mathcal{M}_L}F(\tau)\Es_s(\tau)\dd\mu =& M(F(\tau);s) + \zetas(2s) h_L(s)+\zetas(2s-1) h_L(1-s)\\
			&- \int_{\mathcal{M}-\mathcal{M}_L} \big(R(\tau_2)\left(\zetas(2s)\tau_2^s+\zetas(2s-1) \tau_2^{1-s}\right)+F(\tau)\mathcal{E}_s(\tau)\big)\dd\mu \, .
		\end{split}
	\end{align}
	\noindent		The last term on the right hand side needs further investigation, but the first three terms are easy to deal with. The original Rankin-Selberg method works for the first term since it is integral of a rapid decay function.
	\begin{align}
		\begin{split}
		&\int_{\mathcal{M}-\mathcal{M}_L} \left(R(\tau_2)(\zetas(2s)\tau_2^s+\zetas(2s-1)\tau_2^{1-s})+F(\tau)\mathcal{E}_s(\tau)\right)\dd\mu \\&= \int_L^{\infty}R(\tau_2)e_s(\tau_2)\frac{\dd\tau_2}{\tau_2^2}+\int_L^{\infty}\bigg(\int_0^1 \sum_{m\neq 0}A_m(\tau)\mathcal{E}_s(\tau)\dd\tau_1\bigg)\frac{\dd\tau_2}{\tau_2^2}\\
		& = \int_L^{\infty}R(\tau_2)e_s(\tau_2)\frac{\dd\tau_2}{\tau_2^2}+ \int_L^{\infty}\bigg(\sum_{m\neq 0} \mathcal{E}_{s,m}(\tau_2)a_{-m}(\tau_2)\bigg)\frac{\dd\tau_2}{\tau_2^2} \, ,
		\label{ReInt}
		\end{split}
	\end{align}
	
	\noindent	where we used the notations in equations (\ref{compEis}) to (\ref{GenZeroMode}). Also, we used the fact that
	
\begin{align}
		\int_{\mathcal{M}-\mathcal{M}_L} g(\tau_2) e^{2\pi m i \tau_1} \dd\mu = 0\, ,
\end{align}
	
\noindent	for $m\neq 0$ and a general rapid decay function $g(\tau_2)$. First term of the last line of (\ref{ReInt}) is a rapid decay function of $L$, such as exponential of $L$. So we have

	\begin{align}
		\begin{split}
			\int_{\mathcal{M}_L}F(\tau)\Es_s(\tau)\dd\mu =& M(F;s) + \zetas(2s) h_L(s)+\zetas(2s-1) h_L(1-s)\\
			&- \int_L^{\infty}R(\tau_2)e_s(\tau_2)\frac{\dd\tau_2}{\tau_2^2}+ \int_L^{\infty}\bigg(\sum_{m\neq 0} \mathcal{E}_{s,m}(\tau_2)a_{-m}(\tau_2)\bigg)\frac{\dd\tau_2}{\tau_2^2} \, .
		\end{split}
	\end{align}
}

\noindent Since the residue of $\Es_s(\tau)$ at $s=1$ is $1/2$ we are able to find the integral of $F(\tau)$ over $\mathcal{M}_L$.

\begin{equation}
	\underset{s=1}{\text{Res}} (\zetas(2s) h_L(s)) = \frac{\pi}{6}\alpha_0 \, .
\end{equation}

\begin{equation}
	\underset{s=1}{\text{Res}} (\zetas(2s-1) h_L(1-s)) =\frac{1}{2} \sum_{n\in I \backslash \{1\}} \alpha_n \frac{L^{n-1}}{n-1} +  \frac{\alpha_1}{2} \big(-\gamma_E + \log(4\pi L)\big) \, ,
\end{equation}
\noindent	and also 
\begin{equation}
	\underset{s=1}{\text{Res}}\int_L^{\infty}\bigg(\sum_{m\neq 0} \mathcal{E}_{s,m}(\tau_2)a_{-m}(\tau_2)\bigg)\frac{\dd\tau_2}{\tau_2^2}=0 \, ,
\end{equation}

\noindent because $\mathcal{E}_{m,s}(\tau_2)$ is an entire function of $s$ and the proof of both theorems is done.

\section{Stokes method}

The last method that we discuss in this paper is the Stokes theorem. For this reason, we review the differential operators we will use on modular forms and functions. The first two are Maass operators

\begin{align}
	\begin{split}
		&\nabla = 2 i \tau_2^2 \partial_{\tau} = i\tau_2^2\partial_{\tau_1}+\tau_2^2\partial_{\tau_2} \, ,\\
		&\overline{\nabla} =- 2 i \tau_2^2 \partial_{\bar{\tau}} = -i\tau_2^2\partial_{\tau_1}+\tau_2^2\partial_{\tau_2} \, .
		\label{CRderivatives}
	\end{split}
\end{align}

\noindent Operator $\nabla$ maps a modular form with weight $(0,w^*)$ to a modular form with weight $(0,w^*-2)$ and $\overline\nabla$ maps a modular form with weight $(w,0)$ to a modular form with weight $(w-2,0)$. The combination of these two gives us the Laplace-Beltrami operator, which maps a modular function to a modular function

\begin{equation}
	\label{LapOp}
	\Delta=\nabla \tau_2^{-2} \overline{\nabla}=\overline{\nabla} \tau_2^{-2} \nabla=4 \tau_2^2 \partial_{\bar{\tau}} \partial_\tau=\tau_2^2\left(\partial_{\tau_1}^2+\partial_{\tau_2}^2\right) \, .
\end{equation}

\noindent Then, the Stokes theorem helps us find the integral of the Laplacian of a modular function by only using the derivative of its zero mode over the boundary

	\begin{align}
		\label{Stokes}
		\int_{\mathcal{M}_L} \Delta F(\tau)  \dd \mu  =\left.\int_0^1 \dd \tau_1 \partial_{\tau_2} F(\tau_2)\right|_{\tau_2=L} = \partial_{\tau_2} a_0 \bigg|_{\tau_2=L}\, .
	\end{align}

\noindent The power of this method shows itself when we can relate the Laplacian of a function to itself, in other words, when we have a Laplace (or more precisely, Poisson) equation with a source that we can integrate

\begin{align}
	(\Delta - c) F(\tau) = {\rm source} \, ,
\end{align}

\noindent where $c\neq 0$ and the integral of $F(\tau)$ can be written as

\begin{align}
	\int_{\mathcal{M}_L} F(\tau) \dd\mu = \frac{1}{c} \left(\partial_{\tau_2} a_0 \bigg|_{\tau_2=L}- \int_{\mathcal{M}_L} ({\rm source}) \,\, \dd\mu \right) \, .
\end{align}

In this paper, $c$ is always in the form of $x(x-1)$ for some positive integer $x$, and we denote it by $\mu_x = x(x-1)$. Also, we just consider the equations with non-zero $c$ (which is also called the eigenvalue of the Laplace equation). That is why we want a basis for MGFs with a good Laplace equation. In the next section, we study this basis up to depth three and integrate them.

\section{Depth-dependent basis for MGFs}

The generating function of the modular graph forms can be written as iterated integrals of holomorphic modular forms \cite{Gerken:2020yii, Dorigoni:2022npe}. In this way, we can classify them with the depth of iteration. However, MGFs cannot be graded by depth, and it is just a filtration. The functions appearing in the expansion of the generating function, $\beta^{\rm sv}$ (or the modular form version of them $\beta^{\rm eqv}$) have a nice Cauchy-Riemann equation system. Also, they obey the shuffle algebra. Therefore, each $\beta^{\rm eqv}$ at each depth can be written as the linear combination of the product of the $\beta^{\rm eqv}$'s of lower depth (which we call the \say{shuffle functions}) and some functions that cannot be written as shuffle functions which are called \say{the new functions} or sometimes \say{non-shuffle functions}. Using the Cauchy-Riemann equations, we can choose these new functions to have a nice Laplace equation. In this section, these functions are denoted as $F_{m,k,\cdots}^{\cdots}$, and they are defined by the Laplace equation they satisfy. Table \ref{tableBasis} shows the basis of depth zero to depth three. 
 
\begin{table}[h]
	\begin{center}
	\tikzpicture

	\draw(0.6,1.1) -- (0.6,-4.8);
	\draw(2.2,1.1) -- (2.2,-4.8);
	\draw(4.8,1.1) -- (4.8,-4.8);
	\draw(-1,0.2) -- (8.4,0.2);
	\draw(-1,-0.79) -- (8.4,-0.79);
	\draw(-1,-1.66) -- (8.4,-1.66);
	\draw(-1,-2.4) -- (8.4,-2.4);
	\draw(-1,-3.2) -- (8.4,-3.2);
	\draw(-1,-4) -- (8.4,-4);

	\draw(-0.2,0.5)node{depth 0};
	\draw(1.4,0.5)node{depth 1};
	\draw(3.5,0.5)node{depth 2};
	\draw(6.5,0.5)node{depth 3};
	\draw(-0.17,-0.35)node{1};

	\draw(1.45,-0.35)node{$E_m^*$};

	\draw(3.5,-0.35)node{$\frac{\nabla^{\alpha}E^*_m\overline{\nabla}^{\alpha}E^*_k\pm {\rm c.c.}}{2\tau_2^{2\alpha}}$};
	\draw(3.5,-1.2)node{$F^{*\pm(s)}_{m,k}$};

	\draw(6.7,-0.35)node{$\frac{\nabla^{\alpha}\Es_m\nabla^{\beta}\Es_k\overline\nabla^{\alpha+\beta}\Es_l \pm {\rm c.c.}}{2\tau_2^{2(\alpha+\beta)}}$};
	\draw(6.7,-1.24)node{$\frac{\nabla^{\alpha}\Es_m\overline{\nabla}^{\alpha}F_{k,l}^{*\pm(s)}\pm {\rm c.c.}}{2\tau_2^2}$};
	\draw(6.7,-2)node{$F_{m,k,l}^{*(s)1}$};
	\draw(6.7,-2.8)node{$F_{m,k,l}^{*(w,s)2\pm}$};
	\draw(6.7,-3.6)node{$F_{m,k,l}^{*(s)3\pm}$};
	\draw(6.7,-4.4)node{$F_{m,k,l}^{*(w,s)4\pm}$};

	\endtikzpicture
	\caption{Depth-dependent basis for MGFs up to depth three}
		\label{tableBasis}
\end{center}
\end{table}

\subsection{Depth zero and one}

Depth zero and one are the easiest ones. Depth zero is just a constant. Also, functions of depth one can be written as derivatives of the non-holomorphic Eisenstein series, so the only modular invariant one is $\Es_k(\tau)$ \cite{Dorigoni:2022npe}.

\subsection{Depth two}

The first function in table \ref{tableBasis} is a shuffle function, and it is a product of two depth one functions. The new functions for depth two were explained in great detail for integer weight Eisenstein series in \cite{Dorigoni:2021jfr, Dorigoni:2021ngn} and half-integer weights in \cite{klingerlogan2018differential, fedosova2022whittaker, klingerlogan2022d6}. It is convenient to separate the functions into even and odd under the involution of the upper half-plane $\tau \rightarrow -\bar\tau$. The important note is that this involution does not change $\tau_2$ and therefore, odd functions do not have zero modes, and based on the methods discussed earlier, their integral over $\mathcal{M}_L$ must be exponentially suppressed in $L$. For this reason, we do not write their integral in the next section.

The new functions (without loss of generality, assuming $k \geq m$) at depth two are defined as below

\begin{equation}
	\label{Feven}
	(\Delta-\mu_s) {F}_{m, k}^{*+(s)}=E^*_m E^*_k, \quad s \in\{k-m+2, k-m+4, \ldots, k+m-4, k+m-2\} \, ,
\end{equation}

\noindent where $F^{*+(s)}_{m,k}$ is an even function and again, $\mu_s = s(s-1)$. Note that $s\geq 2$ and therefore $\mu_s \neq 0$. The odd function is also defined as

\begin{equation}
	\begin{aligned}
		(\Delta-\mu_s) {F}_{m, k}^{*-(s)} & =\frac{\nabla \Es_m\overline{\nabla} \Es_k-{\rm c.c.}}{2 \tau_2^2} \\
		s & \in\{k-m+1, k-m+3, \ldots, k+m-3, k+m-1\} \, ,
	\end{aligned}
\end{equation}

\noindent where c.c. stands for complex conjugate. Here, we used $F^*$ rather than $F$, which is used in \cite{Dorigoni:2021jfr,Dorigoni:2021ngn} since we work with $\Es$ instead of $E$. Similar to (\ref{EtoEstar}), the conversion is

\begin{align}
	F_{m,k}^{*\pm (s)} = \frac{\Gamma(m)\Gamma(k)}{4} F_{m,k}^{\pm (s)} \, .
\end{align}

\subsection{Depth three}

The first two rows in table \ref{tableBasis} are shuffle functions. The first is the product of three depth one functions, and the second is the product of a depth one and a depth two new function. After that, we have new functions defined by their Laplace equations. These functions were found in \cite{Dorigoni2023:appear}. We begin by writing some of the examples. At weight six we have

\begin{align}
	(\Delta - 2) F = \frac{1}{6} {\Es_2}^3 + \Es_2 F_{2,2}^{*+(2)} \, ,
	\label{F222}
\end{align}

\noindent at weight seven, we have

\begin{align}
		\Delta F_1 = & {\Es_2}^2 \Es_3 + \frac{1}{3}\left( \frac{\Es_2 \nabla\Es_2\overline\nabla\Es_3+{\rm c.c.}}{2\tau_2^2}\right) - \frac{1}{6} \left( \frac{ \nabla\Es_2\overline\nabla F_{2,3}^{*-(2)}-{\rm c.c.}}{2\tau_2^2}\right) \, , \label{F1Lap}\\
(\Delta - 6)F_2 =& {\Es_2}^2 \Es_3 + 2 \Es_3 F_{2,2}^{*+(2)}+ 12  \Es_2 F_{2,3}^{*+(3)} \, ,\\
 (\Delta - 6)F_3 = &-\frac{1}{2}\left( \frac{\Es_2 \nabla\Es_2\overline\nabla\Es_3+{\rm c.c.}}{2\tau_2^2}\right) + \frac{1}{2} \left( \frac{ \nabla\Es_2\overline\nabla F_{2,3}^{*-(2)}-{\rm c.c.}}{2\tau_2^2}\right)  \\
&+ 3 \Es_3 F_{2,2}^{*+(2)}- 9 \Es_2 F_{2,3}^{*+(3)}\, ,\notag \\
 (\Delta - 6)F_4 = &-\frac{1}{2}\left( \frac{\Es_2 \nabla\Es_2\overline\nabla\Es_3+{\rm c.c.}}{2\tau_2^2}\right) +  \left( \frac{ \nabla\Es_2\overline\nabla F_{2,3}^{*-(2)}-{\rm c.c.}}{2\tau_2^2}\right)  \\
&-2 \Es_3 F_{2,2}^{*+(2)}+16 \Es_2 F_{2,3}^{*+(3)}\, , \notag \\
(\Delta - 20) F_5 = & {\Es_2}^2 \Es_3 -\frac{1}{2}\left( \frac{\Es_2 \nabla\Es_2\overline\nabla\Es_3+{\rm c.c.}}{2\tau_2^2}\right) +\left( \frac{ \nabla\Es_2\overline\nabla F_{2,3}^{*-(2)}-{\rm c.c.}}{2\tau_2^2}\right) \, , 
\label{F22320}
\end{align}

\noindent In this paper, instead of going through all these term, we take each function on the right hand side as a single source, and then we investigate a function that satisfies a Laplace equation with that single source. In this way, the functions above are linear combinations of these new functions, and thus, we call them a basis for depth three non-shuffle functions.

The first one is an even function

\begin{equation}
	(\Delta - \mu_s)F_{m,k,l}^{*(s)1} = E^*_m E^*_k E^*_l \, .
	\label{firstNonshuf}
\end{equation}
The second one has an even and an odd sector. The even sector is denoted by a superscript $+$ and the odd with $-$. Its Laplace equation is

\begin{equation}
	(\Delta - \mu_w)F_{m,k,l}^{*(w,s)2\pm} = E^*_m F_{k,l}^{*\pm(s)}.
\end{equation}
The third one also has even and odd sectors and is defined by the following Laplace equation

\begin{equation}
	(\Delta - \mu_s)F_{m,k,l}^{*(s)3\pm} = \frac{E^*_m\nabla E^*_k\overline{\nabla}E^*_l \pm {\rm c.c.}}{2\tau_2^2} \, .
\end{equation}
And finally, the last one is defined by the following equation, with even and odd sectors

\begin{equation}
	(\Delta - \mu_w)F_{m,k,l}^{*(w,s)4\pm} = \frac{\nabla{E^*_m}{\overline\nabla}F_{k,l}^{*\mp(s)} - {\rm c.c.}}{2\tau_2^2} \, .
	\label{lastnonshuf}
\end{equation}

\noindent The functions in (\ref{F222}) to (\ref{F22320}) up to a homogeneous term, can be written as

\begin{align}
	F =& \,\frac{1}{6} F_{2,2,2}^{*(2)1} + F_{2,2,2}^{*(2,2)2+} \, , \label{FirstLinComb}\\
	F_1 =&\, F_{3,2,2}^{*(1)1}+\frac{1}{3} F_{2,2,3}^{*(1)3+}-\frac{1}{6}F_{2,2,3}^{*(1,2)4+} \, , \label{F1}\\
	F_2 = &\, F_{3,2,2}^{*(3)1} + 2 F_{3,2,2}^{*(3,2)2+}+12 F_{2,2,3}^{*(3,3)2+}  \, ,\\
	F_3 = & \,-\frac{1}{2} F_{2,2,3}^{*(3)3+}+\frac{1}{2}F_{2,2,3}^{*(3,2)4+}+3F_{3,2,2}^{*(3,2)2+}-9F_{2,2,3}^{*(3,3)2+} \, , \\
	F_4 = & \,-\frac{1}{2} F_{2,2,3}^{*(3)3+}+F_{2,2,3}^{*(3,2)4+}-2F_{3,2,2}^{*(3,2)2+}+16F_{2,2,3}^{*(3,3)2+} \, , \\
	F_5 =& \, F_{3,2,2}^{*(5)1}  -\frac{1}{2} F_{2,2,3}^{*(5)3+}+F_{2,2,3}^{*(5,2)4+} \, . \label{LastLinComb}
\end{align}

Although the non-shuffle functions in string theory, as depicted in equations (\ref{F222}) to (\ref{F22320}), are modular functions, it is crucial to recognize that this does not automatically guarantee that equations (\ref{firstNonshuf}) to (\ref{lastnonshuf}) yield modular solutions. The existence of modular solutions is an interesting problem \cite{fedosova2022whittaker, klingerlogan2022d6} but lies beyond the scope of this paper. If a solution is non-modular, Stokes theorem (\ref{Stokes}) introduces a boundary term in addition to the desired terms; however, by summing various terms, as demonstrated in equations (\ref{FirstLinComb}) to (\ref{LastLinComb}), all these boundary terms cancel out since these linear combinations of the basis are modular. Therefore, even though our integrations include a potential boundary term (which is not determined in this paper), applying this basis to modular graph functions does not present an issue. So when we write \say{=} in the integral of this basis, we mod out the boundary terms.

An important difference between depth three and two is that at depth three, one can have a Laplace equation with a vanishing eigenvalue, \textit{i.e.} $\Delta F =$ source. An instance in weight seven is equation (\ref{F1Lap}). This function cannot be integrated by Stokes theorem. Furthermore, the homogeneous answer to this equation is a constant, which contributes to the integral over $\mathcal{M}_L$. There is no way to fix this constant by just using the Laplace equation and imposing modularity. To find this constant, we must write this function in terms of $\beta^{\rm eqv}$, and the result is given in \cite{Dorigoni2023:appear}. Thus, the integral of this function requires the integration of $\beta^{\rm eqv}$'s, which will be discussed in the future \cite{Doroudiani2023appear}.

All the other even functions up to weight seven have non-vanishing eigenvalues; therefore, we can use the Stokes theorem to integrate them. 

One last note before we study the integrals is that the shuffle functions are products of the derivatives of lower depth functions. However, if the number of derivations is too high, they do not have any contribution to MGFs. Therefore, the number of derivations must be in the following spectrum

\begin{align}
	\label{spec}
	\begin{split}
		&\nabla^{\alpha} E^*_m : \hspace{3cm} \alpha \in \{0,1,\cdots, m-1\}\, ,\\
		&\overline\nabla^{\beta} E^*_m : \hspace{3cm} \beta \in \{0,1,\cdots, m-1\}\, ,\\
		&\nabla^{\gamma} {F}_{m, k}^{*\pm(s)}  : \hspace{2.5cm} \gamma \in \{0,1,\cdots, k+m-2\}\, ,\\
		&\overline\nabla^{\delta} {F}_{m, k}^{*\pm(s)}  : \hspace{2.55cm} \delta \in \{0,1,\cdots, k+m-2\}\, .
	\end{split}
\end{align}
Furthermore, for $\gamma, \delta \geq s$ functions $\nabla^{\gamma} {F}_{m, k}^{*\pm(s)}$ and $\overline\nabla^{\delta} {F}_{m, k}^{*\pm(s)}$ become shuffle functions \cite{Dorigoni:2021jfr,Dorigoni:2021ngn} and therefore they would be redundant to the other shuffle functions in table \ref{tableBasis}. Nevertheless, the calculations presented here are general for any number of derivations in the shuffle functions.

\section{Integrating the basis over $\mathcal{M}_L$}

To report the results of the integrals, it is essential to mention that all the $L$-dependent terms in the results must cancel out when we compute the full amplitude since it is just a cut-off to partition the moduli space. Although from \cite{DHoker:2019blr}, we can see that the coefficient of the logarithmic terms of $L$ in the integral over $\mathcal{M}_R$ is the same as the coefficient of the logarithmic terms of the kinematic variables. For this reason, keeping the logarithmic term of the integral is useful. In addition, theorem \hyperlink{Th3}{3} is only valid for large ${\rm Re}(s)$, so when we use it, we must analytically continue the result to other values of $s$. In this case, terms that are not exponentially suppressed in $L$ could give us constant or logarithmic terms. So from now on, whenever we use \say{$=$}, we mod out all the exponentially suppressed terms since they have no contribution in the full integral over the fundamental domain. Also, when we use \say{$\approx$}, we mod out all the $L$-dependent terms except the logarithmic one. The final results of the integrals of even functions are as follows

\begin{equation}
	\int_{\mathcal{M}_L} \dd\mu \approx \frac{\pi}{3} \, ,
\end{equation}

\begin{equation}
	\int_{\mathcal{M}_L}  \Es_m(\tau) \dd\mu  \approx 0 \, ,
\end{equation}

\begin{align}
	\int_{\mathcal{M}_L}  &\tfrac{ \nabla^{\alpha}E^*_m \overline\nabla^{\alpha}E^*_k + {\rm c.c}}{2\tau_2^{2\alpha}} \,  \dd\mu \approx\, 2(m)_{\alpha}(1-m)_{\alpha}\zetas(2m)\zetas(2m-1)\, \delta_{m,k}   \notag \\
	&\times\bigg(\log L+ \frac{{\zetas}'(2m)}{\zetas(2m)}-\frac{{\zetas}'(2m-1)}{\zetas(2m-1)}+\frac{1}{2} \left(H_1(m+\alpha)-2 H_1(m)+H_1(m-\alpha)\right)\bigg) \, ,
\end{align}

\noindent where $(x)_n$ is the ascending Pochhammer symbol

\begin{align}
	(x)_{\alpha} = x(x+1)\cdots(x+\alpha-1) = \frac{\Gamma(x+\alpha)}{\Gamma(x)} \, ,
\end{align}
and
\begin{align}
	H_1(x) = \sum_{n=1}^{x-1} \frac{1}{n} \, ,
\end{align}
with $H_1(1)=1$ is a harmonic sum. 

\noindent The integral of the next one is

\begin{equation}
	\int_{\mathcal{M}_L}F_{m,k}^{*+(s)}\dd\mu \approx	-\frac{1}{\mu_s} \delta_{m,k} \left(\int_{\mathcal{M}_L} \Es_m\Es_k \dd\mu + \frac{2\zetas(2m)\zetas(2m-1)}{\mu_s} \right) \, .
\end{equation}

The next function is $\tfrac{\nabla^{\alpha} E_m^* \nabla^{\beta} E_k^* \overline\nabla^{\alpha+\beta} E_l^* + {\rm c.c.}}{2\tau_2^{2(\alpha+\beta)}}$ and its integral is more complicated. The result will be explained in equations (\ref{EEEint}) to (\ref{EEEintFinal}). For the next ones we have

\begin{align}
	\int_{\mathcal{M}_L} &\tfrac{\nabla^{\alpha} E^*_m\overline\nabla^{\alpha}F^{*+(m)}_{k,l} + {\rm c.c.}}{2\tau_2^{2\alpha}}  \dd\mu \approx \frac{(m)_{\alpha}(1-m)_{\alpha}}{2m-1}\notag \\
	& \times\zetas(m+k+l-1)\zetas(m-k+l)\zetas(m+k-l)\zetas(-m+k+l) \notag\\
	&\times \bigg(\frac{-1}{2m-1}+H_1(m-\alpha)-H_1(m) - \log L -2\frac{{\zetas}'(2m)}{\zetas(2m)}-\frac{{\zetas}'(-m+k+l)}{\zetas(-m+k+l)}\notag\\
	&\hspace{20pt} +\frac{{\zetas}'(m+k+l-1)}{\zetas(m+k+l-1)}+\frac{{\zetas}'(m-k+l)}{\zetas(m-k+l)}+\frac{{\zetas}'(m+k-l)}{\zetas(m+k-l)}\bigg) \, ,
\end{align}
and for $m\neq s$ we have

\begin{align}
	\int_{\mathcal{M}_L} &\tfrac{\nabla^{\alpha} E^*_m\overline\nabla^{\alpha}F^{*+(s)}_{k,l} + {\rm c.c.}}{2\tau_2^{2\alpha}}  \dd\mu \approx \frac{(m)_{\alpha}(1-m)_{\alpha}}{\mu_m-\mu_s}\bigg( \int_{\mathcal{M}_L} \Es_m \Es_k \Es_l \dd\mu \notag \\
	&+ \sum_{\mathfrak{m},\mathfrak{k},\mathfrak{l}} \zetas(2\mathfrak{m})\zetas(2\mathfrak{k})\zetas(2\mathfrak{l}) \bigg(\frac{\mathfrak{m}-1}{\mu_m-\mu_s} + H_1(1-\mathfrak{m}+\alpha)-H_1(1-\mathfrak{m})\bigg) \delta_{\mathfrak{m+k+l},1}\bigg) \, .
\end{align}
In this notation, gothic font $\mathfrak{m}, \mathfrak{k}$ and $\mathfrak{l}$ mean that they are chosen from the set $\{m,1-m\}$, $\{k,1-k\}$ and $\{l, 1-l\}$ respectively. We will use this font with the same meaning throughout this paper.

\begin{align}
	\int_{\mathcal{M}_L}	\frac{\nabla^{\alpha}\Es_m\overline\nabla^{\alpha} F_{k,l}^{*-(s)}-{\rm c.c.}}{2\tau_2^{2\alpha}} \dd\mu  \approx 0 \, ,
\end{align}

\begin{align}
	\int_{\mathcal{M}_L}F_{m,k,l}^{*(s)1}\dd\mu
	\approx & -\frac{1}{\mu_s} 	\left(\int_{\mathcal{M}_L}  \Es_{m}\Es_{k}\Es_{l} \dd\mu +\frac{1}{\mu_s}\sum_{\mathfrak{m},\mathfrak{k},\mathfrak{l}} {\zetas(2\mathfrak{m})\zetas(2\mathfrak{k})\zetas(2\mathfrak{l})}\delta_{\mathfrak{m}+\mathfrak{k}+\mathfrak{l},1} \right)\, ,
\end{align}

\begin{align}
	\int_{\mathcal{M}_L}F_{m,k,l}^{*(w,m)2+} \dd\mu\approx &\, \tfrac{-1}{\mu_w(2m-1)}\zetas(m+k+l-1)\zetas(m-k+l)\zetas(m+k-l)\zetas(-m+k+l) \notag\\
	&\times \bigg(\frac{2m-1-\mu_w}{\mu_w(2m-1)} - \log L -2\frac{{\zetas}'(2m)}{\zetas(2m)}-\frac{{\zetas}'(-m+k+l)}{\zetas(-m+k+l)}\notag\\
	&+\frac{{\zetas}'(m+k+l-1)}{\zetas(m+k+l-1)}+\frac{{\zetas}'(m-k+l)}{\zetas(m-k+l)}+\frac{{\zetas}'(m+k-l)}{\zetas(m+k-l)}\bigg) \, .
\end{align}

Also, for $m\neq s$ we have

\begin{align}
	\int_{\mathcal{M}_L}F_{m,k,l}^{*(w,s)2+} \dd\mu\approx &\, \tfrac{-1}{\mu_w(\mu_m-\mu_s)}\bigg( \int_{\mathcal{M}_L} \Es_m \Es_k \Es_l \dd\mu -\tfrac{1}{\mu_w(\mu_m-\mu_s)} \sum_{\mathfrak{m},\mathfrak{k},\mathfrak{l}}\zetas(2\mathfrak{m}) \zetas(2\mathfrak{k}) \zetas(2\mathfrak{l}) \notag \\
	&\times \bigg(\mu_m-\mu_s-\mu_w(2\mathfrak{m}-1)\bigg) \delta_{\mathfrak{m+k+l},1} \bigg) \, ,
\end{align}

\begin{align}
	\int_{\mathcal{M}_L}F_{m,k,l}^{*(s)3+} \dd\mu  \approx  &\frac{-\mu_m+\mu_k+\mu_l}{2\mu_s} \int_{\mathcal{M}_L} \Es_m \Es_k \Es_l \dd\mu \notag \\
	&+	\sum_{\mathfrak{m},\mathfrak{k},\mathfrak{l}}\zetas(2\mathfrak{m})\zetas(2\mathfrak{k})\zetas(2\mathfrak{l}) \bigg(\frac{-\mu_m+\mu_k+\mu_l}{2\mu_s^2}+\frac{2\mathfrak{m}-1}{2\mu_s}\bigg)\delta_{\mathfrak{m+k+l},1} \, ,
\end{align}

\begin{align}
		\int_{\mathcal{M}_L}F_{m,k,l}^{*(w,s)4+} \dd\mu \approx 0 \, .
\end{align}

\noindent Now, we prove these results in the following subsections.

\subsection{Depth zero}

Even though depth zero can be integrated directly without difficulty, it is a good place to practice theorem \hyperlink{Th2}{2} with it. For a constant function we just have $\alpha_0=1$ and all the other $\alpha$'s are zero. {$R(\tau_2)$ and $a_m$'s are also zero.} Therefore, using equation (\ref{ThTwo}) we have

\begin{equation}
	\int_{\mathcal{M}_L} \dd\mu = \frac{\pi}{3} - \frac{1}{L} \approx \frac{\pi}{3} \, .
\end{equation}

\subsection{Depth one}

The easiest way to integrate the non-holomorphic Eisenstein series is to use its Laplace equation

\begin{align}
	(\Delta - \mu_m) \Es_m = 0 \, .
\end{align}

\noindent Again, we take a more difficult integration route to test our other theorems. By theorem \hyperlink{Th2}{2}, we have $\alpha_m = \zetas(2m)$ and $\alpha_{1-m}=\zetas(2m-1)$ (note that $m\neq 1, 0$), $R(\tau_2)$ is zero and $a_k(\tau_2)=\mathcal{E}_{m,k}(\tau_2)$. Then we have

\begin{equation}
	\label{IntEis}
	\int_{\mathcal{M}_L}  \Es_m(\tau) \dd\mu = \zetas(2m)\frac{L^{m-1}}{m-1}-\zetas(2m-1)\frac{L^{-m}}{m} = \sum_{\mathfrak{m}} \zetas(2\mathfrak{m}) \frac{L^{\mathfrak{m}-1}}{\mathfrak{m}-1}\, \approx 0 \, ,
\end{equation}

\noindent where again, we used the gothic notation $\mathfrak{m} = \{m,1-m\}$. This integral was first calculated in \cite{zagier1981rankin} to regularize the integral of the non-holomorphic Eisenstein series over $\mathcal{M}$. This is also the same result we get from theorem \hyperlink{Th3}{3} with the constant function 1. As equation (\ref{IntEis}) indicates, depth one has no contribution to the string amplitude.

\subsection{Depth two}

\subsubsection{Shuffle functions}

Before integrating the depth two basis, it is useful to focus on $\Es_m\Es_k$. As we will see in the following subsection, this problem is the essence of the other depth two basis. Once again, the easiest way is to use the Laplace equation, but instead, we use theorems \hyperlink{Th2}{2} and \hyperlink{Th3}{3}. This process is helpful for depth three as well.

The zero mode of $\Es_m\Es_k$ has a Laurent polynomial and a non-perturbative part

\begin{align}
	\label{zeroModeEE}
	a_0(\tau_2) = \, &\zetas(2m)\zetas(2k)\tau_2^{m+k}+ \zetas(2m)\zetas(2k-1)\tau_2^{1+m-k} \notag \\
&	+ \zetas(2m-1)\zetas(2k)\tau_2^{1-m+k} + \zetas(2m-1)\zetas(2k-1)\tau_2^{2-m-k}+R(\tau_2) \, , \notag\\
=& \sum_{\mathfrak{m},\mathfrak{k}} \zetas(2\mathfrak{m})\zetas(2\mathfrak{k}) \tau_2^{\mathfrak{m}+\mathfrak{k}} + R(\tau_2)
\end{align}

\noindent where $R(\tau_2)$ is

\begin{equation}
	R(\tau_2)=8\tau_2\sum_{N=1}^{\infty} N^{m+k-1}\sigma_{1-2m}(N)\sigma_{1-2k}(N)K_{m-1/2}(2\pi N\tau_2)K_{k-1/2}(2\pi N\tau_2) \, .
\end{equation}

\noindent The Mellin transform of the non-perturbative part is

\begin{align}
	\begin{split}
		M(\Es_m\Es_k;s) = 8\zetas(2s) &\bigg(\sum_{N=1}^{\infty}N^{m+k-1}\sigma_{1-2m}(N)\sigma_{1-2k}(N)\\
	\times&	\int_0^{\infty}\tau_2^{s-1}K_{m-1/2}(2\pi N\tau_2)K_{k-1/2}(2\pi N\tau_2)\dd\tau_2 \bigg)\\
		=\frac{8\zetas(2s)}{(2\pi)^{s}}&\bigg(\sum_{N=1}^{\infty}N^{m+k-s-1}\sigma_{1-2m}(N)\sigma_{1-2k}(N)\bigg) \\
		\times &\bigg(\int_0^{\infty}u^{s-1}K_{m-1/2}(u)K_{k-1/2}(u) \dd u\bigg) \, .
	\end{split}
\end{align}

\noindent Here we have a Ramanujan summation and from \cite{zagier1981rankin} we have

\begin{align}
	\begin{split}
		\sum_{N=1}^{\infty}&N^{m+k-s-1}\sigma_{1-2m}(N)\sigma_{1-2k}(N)=\\
		&\frac{1}{\zeta(2s)}\zeta(s+m+k-1)\zeta(s-m+k)\zeta(s+m-k)\zeta(s-m-k+1) \, ,
	\end{split}
	\label{RamSum}
\end{align}

\noindent and from \cite{erdelyi1954tables}, the integral is

\begin{align}
	\begin{split}
		\label{eqIntBessel}
		\int_0^{\infty}&u^{s-1}K_{m-1/2}(u)K_{k-1/2}(u)\dd u =\\ &\frac{2^{s-3}}{\Gamma(s)}\Gamma\left(\frac{s+m+k-1}{2}\right)\Gamma\left(\frac{s-m+k}{2}\right)\Gamma\left(\frac{s+m-k}{2}\right)\Gamma\left(\frac{s-m-k+1}{2}\right) \, .
	\end{split}
\end{align}

\noindent By putting them together, we have

\begin{equation}
	M(\Es_m\Es_{k};s) = \zeta^*(s+m+k-1)\zeta^*(s-m+k)\zeta^*(s+m-k)\zeta^*(s-m-k+1) \, ,
\end{equation}

\noindent which is the result first proven in \cite{zagier1981rankin}. We must study the pole structure of $M(\Es_m\Es_{k};s)$. For this, it is good to write the expansion of $\zetas(s)$ near zero (from the functional relation, we can also conclude it around one)

\begin{equation}
	\zetas(s) = -\frac{1}{s} + \frac{1}{2} (\gamma_E - \log (4\pi)) + \mathcal{O}(s) \, .
\end{equation}

\noindent Therefore, the function $M(\Es_m\Es_{k};s)$ has simple poles at $s=1$ for $m=k$ and $m=k\pm1$, otherwise $M(\Es_m\Es_{k})$ is zero. In the case of $M(\Es_m\Es_{k})=0$, we do not have any constant or linear term in the Laurent polynomial, and we can write

\begin{align}
	\label{doubleEis}
	\int_{\mathcal{M}_L}  \Es_m\Es_{k} \dd\mu =\, &\zetas(2m)\zetas(2k)\frac{L^{m+k-1}}{m+k-1}+ \zetas(2m)\zetas(2k-1)\frac{L^{m-k}}{m-k} \notag \\
	&+\zetas(2m-1)\zetas(2k)\frac{L^{-m+k}}{-m+k}+ \zetas(2m-1)\zetas(2k-1)\frac{L^{1-m-k}}{1-m-k} \notag \\
	=& \sum_{\mathfrak{m},\mathfrak{k}} \zetas(2\mathfrak{m})\zetas(2\mathfrak{k})\frac{L^{\mathfrak{m}+\mathfrak{k}-1}}{\mathfrak{m}+\mathfrak{k}-1} \, ,
\end{align}

\noindent where again, the gothic fonts $\mathfrak{m}$ and $\mathfrak{k}$ mean $\mathfrak{m}= \{m,1-m\}$ and $\mathfrak{k}= \{k,1-k\}$.

In the case of $m=k\pm 1$, we would have a constant term in the Laurent polynomial that cancels the contribution of $M(\Es_m\Es_{k})$ according to equation (\ref{ThTwo}) and we get the same answer as equation (\ref{doubleEis}).

We can also get this result from theorem \hyperlink{Th3}{3} by taking $F(\tau) = \Es_k$. Then $R(\tau) = 0$. Note that since theorem \hyperlink{Th3}{3} only works for large ${\Re}(m)$, we cannot directly get the result of $m=k$. For that, we must analytically continue the result and take the limit $m \rightarrow k$. This limit was taken in \cite{Green:2008uj} and \cite{DHoker:2019mib}.

For $m= k$, there is a linear term in the Laurent polynomial. Adding all the terms of equation (\ref{ThTwo}) gives us

\begin{align}
	\begin{split}
		\int_{\mathcal{M}_L}  \Es_m(\tau)^2 \dd\mu =\, &\zetas(2m)^2\frac{L^{2m-1}}{2m-1}+ \zetas(2m-1)^2\frac{L^{1-2m}}{1-2m}+2\zetas(2m)\zetas(2m-1) \log L \\
		&+ 2\zetas(2m-1) \frac{\dd \zetas(2m)}{\dd m}  - 2\zetas(2m) \frac{\dd \zetas(2m-1)}{\dd m} \, .
	\end{split}
\end{align}

\noindent Then by modding out the $L$-dependent terms (except the logarithmic one), we have

\begin{equation}
	\int_{\mathcal{M}_L} E_m^*(\tau)E_{k}^*(\tau) \dd\mu \approx 2\zetas(2m)\zetas(2m-1)\left(\log L + \frac{{\zetas}'(2m)}{\zetas(2m)}-\frac{{\zetas}'(2m-1)}{\zetas(2m-1)}\right) \delta_{m,k} \, ,
\end{equation}

\noindent where

\begin{align}
	 {\zetas}'(x_0) = \frac{\dd \zetas(x)}{\dd x} \bigg|_{x=x_0} \, .
\end{align}

We can use the following formulas to convert $\zetas$ and its derivative to normal $\zeta$ and other transcendental numbers. For a positive integer $x$ we have

\begin{align}
	&\zetas(2x) = \zetas(-2x+1) = (-1)^{x+1}(2\pi)^x\frac{(2x-2)!!}{(2x)!}B_{2x} \, ,\\
	&\zetas(2x+1) = \zetas(-2x) = \frac{(2x-1)!!}{(2\pi)^x} \zeta(2x+1) \, ,
\end{align}
where $B_{2x}$ is the Bernoulli number and the double factorial is defined by 

\begin{equation}
	x!!=\prod_{k=0}^{\left\lceil\frac{x}{2}\right\rceil-1}(x-2 k)=x(x-2)(x-4) \cdots \, .
\end{equation}
Furthermore, we used the following relation for even zeta values

\begin{align}
	\zeta(2x) = (-1)^{x+1} \frac{(2\pi)^{2x} }{2(2x)!}B_{2x} \, .
\end{align}
Also, for the derivative we have

\begin{align}
	\frac{{\zetas}'(x)}{\zetas(x)} = - \frac{{\zetas}'(-x+1)}{\zetas(-x+1)} = \frac{\zeta'(x)}{\zeta(x)} + \frac{1}{2}\psi\left(x/2\right) - \frac{1}{2} \log\pi
\end{align}
where $\psi(x) = \Gamma^{\prime}(x)/\Gamma(x)$ is digamma. We can convert digamma to harmonic numbers.

\begin{align}
	\label{psiToH}
	\begin{split}
	&\psi(x) = H_1(x) - \gamma_E \, , \\
    &\psi(x+1) = 2H_1(2x+1) - H_1(x+1) - 2 \log 2 - \gamma_E \, .
	\end{split}
\end{align}
Then for even and odd positive numbers, we have

\begin{align}
	\frac{{\zetas}'(2x)}{\zetas(2x)}&= \frac{{\zeta}'(2x)}{\zeta(2x)} +\frac{1}{2}H_1(x)-\frac{1}{2}\left(\gamma_E+\log\pi\right) \\
	\frac{{\zetas}'(2x-1)}{\zetas(2x-1)}&= \frac{{\zeta}'(2x-1)}{\zeta(2x-1)} +H_1(2x+1) - \frac{1}{2} H_1(x+1) -\frac{1}{2}\left(\gamma_E+\log4\pi\right) \, .
\end{align}
By using this dictionary, it is easy to convert $\zetas$ to more familiar objects but for the rest of this work, we use $\zetas$ and its derivatives.

{$\blacksquare$}  Now we go through table \ref{tableBasis}. The first function is a shuffle of two depth one functions. As discussed before, we do not need to consider the odd functions. So we just look at

 \begin{equation}
	G_{m,k}^{\alpha} = \frac{ \nabla^{\alpha}E^*_m \overline\nabla^{\alpha}E^*_k + {\rm c.c}}{2\tau_2^{2\alpha}} \, .
\end{equation}

 \noindent To find the integral, we use theorem \hyperlink{Th1}{1} and we assume that $m > k$ to have a convergent Poincar\'e sum, and then we unfold $\nabla^{\alpha}\Es_m$. Therefore, the zero mode of the seed function is

\begin{align}
	f_0 = (m)_{\alpha} \zetas(2&m)\tau_2^{m+\alpha}\bigg(\zetas(2k)(k)_{\alpha}\tau_2^{k-\alpha} + \zetas(2k-1)(1-k)_{\alpha} \tau_2^{1-k-\alpha}\bigg) \, .
\end{align}

\noindent By using the unfolding method (\ref{ThOne}), we get the following result

\begin{equation}
	\int_{\mathcal{M}_L} G_{m,k}^{\alpha} \dd\mu = \sum_{\mathfrak{m},\mathfrak{k}} \zetas(2\mathfrak{m})\zetas(2\mathfrak{k})(\mathfrak{m})_{\alpha}
	(\mathfrak{k})_{\alpha} \frac{L^{\mathfrak{m}+\mathfrak{k}-1}}{\mathfrak{m}+\mathfrak{k}-1} \, .
\end{equation}
Now we use three identities take the limit $m\rightarrow k $ where this integral results in a constant and logarithmic value. First we use 
\begin{align}
	\label{pochref}
	(-x)_{\alpha} = (-1)^{\alpha}(x-\alpha +1)_{\alpha} \, ,
\end{align}
to invert negative arguments of Pochhammer to positive ones. Next, we need the derivation of Pochhammer (with positive arguments) to take the limit
\begin{align}
	\label{pochdif}
	\frac{\dd (x)_{\alpha}}{\dd x} = (x)_{\alpha} \left(\psi(x+\alpha)-\psi(x)\right)= (x)_{\alpha} \left(H_1(x+\alpha)-H_1(x)\right) \, ,
\end{align}
 By using (\ref{psiToH}), we have

\begin{align}
\int_{\mathcal{M}_L} G_{m,k}^{\alpha} \dd\mu \approx &\, (m)_{\alpha}(1-m)_{\alpha}\, \delta_{m,k} \bigg(\int_{\mathcal{M}_L} E^*_m E^*_k \dd\mu+\zetas(2m)\zetas(2m-1) \notag \\
&\times \left(H_1(m+\alpha)-2 H_1(m)+H_1(m-\alpha)\right)\bigg) \, .
\end{align}

\noindent One can get the same result using the recursive equations from the Laplace equation of these shuffle functions. For $\alpha\geq1$ we have

\begin{equation}
	\left(\Delta-(\mu_m+\mu_k-2\alpha^2)\right)G_{m,k}^{\alpha} = G_{m,k}^{\alpha+1} + (\mu_m-\mu_{\alpha})(\mu_k-\mu_{\alpha})G_{m,k}^{\alpha-1} \, .
\end{equation}

\subsubsection{Non-shuffle functions}

{$\blacksquare$} The next even function in the list is $F_{m,k}^{*+(s)}$. Using (\ref{Feven}) and (\ref{Stokes}) we have

\begin{align}
	\label{IntFeven}
	\begin{split}
		\int_{\mathcal{M}_L}F_{m,k}^{*+(s)} \dd \mu = \frac{1}{\mu_s} \bigg(\partial_{\tau_2}a_0\bigg|_{\tau_2=L} - \int_{\mathcal{M}_L}{E^*_mE^*_k}\dd\mu \bigg) \, ,
	\end{split}
\end{align}

\noindent where $a_0$ is the zero mode of $F_{mk}^{*+(s)}$. Finding the zero mode using the Laplace equation is easy, although we cannot get the homogeneous solution. The new functions need homogeneous terms to be modular invariant. The homogeneous term of $F_{m,k}^{+(s)}$ is written in \cite{Dorigoni:2021jfr,Dorigoni:2022bcx,Green:2008bf}. All of the homogeneous terms of the Laplace equation are monomials proportional to $\tau_2^s$ or $\tau_2^{1-s}$ and for the depth two new functions, the coefficient of $\tau_2^s$ is zero. The zero mode of $F_{m,k}^{*+(s)} $ is 

\begin{align}
	\begin{split}
    a_0 &=\sum_{\mathfrak{m},\mathfrak{k}} \zetas(2\mathfrak{m})\zetas(2\mathfrak{k})\frac{\tau_2^{\mathfrak{m}+\mathfrak{k}}}{\mu_{\mathfrak{m}+\mathfrak{k}}-\mu_{s}} + c_{m,k}^{(s)} (1-s) \tau_2^{1-s}\, ,
	\end{split}
\end{align}
\noindent where the last term is the homogeneous term and the coefficient of $c_{m,k}^{(s)}$ can be found from equations (3.8) and (3.10) of \cite{Dorigoni:2022bcx}
\begin{align}
	\label{hom}
	c_{m,k}^{(s)} = \frac{-\zetas(s+m+k-1)\zetas(s-m+k)\zetas(s+m-k)\zetas(-s+m+k)}{(2s-1)\zetas(2s)} \, .
\end{align}
 Combining this with (\ref{IntFeven}) and (\ref{doubleEis}) gives us the result for $m\neq k$

\begin{align}
	\begin{split}
		\int_{\mathcal{M}_L}F_{m,k}^{*+(s)}\dd\mu =  \sum_{\mathfrak{m},\mathfrak{k}} \zetas(2\mathfrak{m})\zetas(2\mathfrak{k})\frac{L^{\mathfrak{m}+\mathfrak{k}-1}}{(\mathfrak{m}+\mathfrak{k}-1)(\mu_{\mathfrak{m}+\mathfrak{k}}-\mu_{s})}+  c_{m,k}^{(s)} (1-s) L^{-s} \, .
	\end{split}
\end{align}

\noindent For $m=k$ we can take the limit of this equation. The final result is

\begin{equation}
	\int_{\mathcal{M}_L}F_{m,k}^{*+(s)}\dd\mu \approx	-\frac{1}{\mu_s} \delta_{m,k} \left(\int_{\mathcal{M}_L} \Es_m\Es_k \dd\mu + \frac{2\zetas(2m)\zetas(2m-1)}{\mu_s} \right) \, .
\end{equation}

\subsection{Depth three}

\subsubsection{Shuffle functions}

Once again, we integrate the simplest case, $\Es_m\Es_k\Es_l$ first. The only theorem that works for this case is theorem \hyperlink{Th3}{3} and taking ${\rm Re}(m)$ to be very large and then analytically continuing it to any other number. This calculation was first done in \cite{zagier1981rankin} and is almost the same as what we did for depth two, using theorem \hyperlink{Th2}{2} so we do not repeat it. The result is

\begin{align}
	\label{tripleEis}
	\int_{\mathcal{M}_L}  \Es_{m}\Es_{k}\Es_{l} \dd\mu = & \, \zeta^*(m+k+l-1)\zeta^*(m-k+l)\zeta^*(m+k-l)\zeta^*(-m+k+l) \notag  \\
	& + \sum_{\mathfrak{m},\mathfrak{k},\mathfrak{l}} \zetas(2\mathfrak{m})\zetas(2\mathfrak{k})\zetas(2\mathfrak{l})\frac{L^{\mathfrak{m}+\mathfrak{k}+\mathfrak{l}-1}}{\mathfrak{m}+\mathfrak{k}+\mathfrak{l}-1} \, .
\end{align}

\noindent The analysis of taking the limit of this equation is discussed in great detail in \cite{DHoker:2021ous}, and the transcendentality properties were found there. Assuming $m \geq k \geq l$, we only have simple poles in two terms when $m = k+l$ or $m = k+l -1$. In both cases, the poles cancel out, and we get a logarithmic term with the derivatives of completed zetas. When $m \neq k+l$ or $k+l-1$ we have

\begin{align}
	\label{tripE1}
	\int_{\mathcal{M}_L}  \Es_{m}\Es_{k}\Es_{l} \dd\mu \approx& \, \zeta^*(m+k+l-1)\zeta^*(m-k+l)\zeta^*(m+k-l)\zeta^*(-m+k+l) \, .
\end{align}

\noindent When $m = k+l$ we have

\begin{align}
	\label{tripE2}
	\int_{\mathcal{M}_L}  \Es_{m}\Es_{k}\Es_{l}\bigg|_{m=k+l} \dd\mu \approx \, &  \zeta^*(2m-1)\zeta^*(2k)\zeta^*(2l)\bigg(\frac{1}{2}(\gamma_E-\log(4\pi)) \notag \\
	& -\frac{{\zetas}'(2m-1)}{\zetas(2m-1)}+\frac{{\zetas}'(2k)}{\zetas(2k)}+\frac{{\zetas}'(2l)}{\zetas(2l)}+ \log L\bigg) \, ,
\end{align}

\noindent and finally, when $m = k+l-1$ we have

\begin{align}
	\label{tripE3}
	\int_{\mathcal{M}_L}  \Es_{m}\Es_{k}\Es_{l}\bigg|_{m=k+l-1} \dd\mu \approx \, &  \zeta^*(2m)\zeta^*(2k-1)\zeta^*(2l-1)\bigg(\frac{1}{2}(\gamma_E-\log(4\pi)) \notag \\
	& +\frac{{\zetas}'(2m)}{\zetas(2m)}-\frac{{\zetas}'(2k-1)}{\zetas(2k-1)}-\frac{{\zetas}'(2l-1)}{\zetas(2l-1)}+ \log L\bigg) \, .
\end{align}

{$\blacksquare$} Now, we can look at table \ref{tableBasis} and integrate the even functions. We start with the product of three depth one functions, which is a shuffle function

\begin{equation}
	\frac{\nabla^{\alpha} E_m^* \nabla^{\beta} E_k^* \overline\nabla^{\alpha+\beta} E_l^* + \rm{c.c.}}{2\tau_2^{2(\alpha+\beta)}} \, .
\end{equation}

\noindent We take $\alpha$ and $\beta$ to be non-negative, and without loss of generality, we unfold the first term in the product. If the Poincar\'e sum is divergent, we shift $m$ to be a large number and use analytic continuation. Since the final result is finite, we do not do this procedure explicitly.

\begin{align}
	f_0 =& \,\zetas(2m) (m)_{\alpha}\tau_2^{m+\alpha}\bigg(\sum_{\mathfrak{k},\mathfrak{l}} \zetas(2\mathfrak{k})\zetas(2\mathfrak{l})(\mathfrak{k})_{\beta}(\mathfrak{l})_{\alpha+\beta} \tau_2^{\mathfrak{k}+\mathfrak{l}-\alpha}
	+ \frac{1}{\tau_2^{2(\alpha+\beta)} }R(\tau_2) \bigg) \, ,
\end{align}

\noindent where $R$ is the rapid decay term in $\nabla^{\beta} E_k^*\overline\nabla^{\alpha+\beta} E_l^*$.  Then the integral is

\begin{align}
	\int_{\mathcal{M}_L} 	\tfrac{\nabla^{\alpha} E_m^* \nabla^{\beta} E_k^* \overline\nabla^{\alpha+\beta} E_l^* + {\rm c.c.}}{2\tau_2^{2(\alpha+\beta)}}  \dd\mu =&  \bigg(\sum_{\mathfrak{m},\mathfrak{k},\mathfrak{l}} \zetas(2\mathfrak{m})\zetas(2\mathfrak{k})\zetas(2\mathfrak{l})(\mathfrak{m})_{\alpha}(\mathfrak{k})_{\beta}(\mathfrak{l})_{\alpha+\beta} \tfrac{L^{\mathfrak{m}+\mathfrak{k}+\mathfrak{l}-1}}{\mathfrak{m}+\mathfrak{k}+\mathfrak{l}-1}\bigg) \notag \\
	&+ \int_0^{\infty} \zetas(2m)(m)_{\alpha} \tau_2^{m-\alpha-2\beta-2} R(\tau_2) \, \, \dd \tau_2 \, .
\end{align}

\noindent To find $R(\tau_2)$, we start with writing the non-zero mode of the Eisenstein series. 

\begin{align}
	\begin{split}
		\mathcal{E}_k&=2\sqrt{\tau_2}\sum_{N\neq0}|N|^{k-1/2}\sigma_{1-2k}(N)K_{k-1/2}(2\pi |N|\tau_2)e^{2\pi iN\tau_1}\\
		&= \sum_{N\neq 0} \mathcal{N}_{N,k} g_{N,k}(\tau_2) e^{2\pi i N\tau_1} \, ,
	\end{split}
\end{align}

\noindent where we introduced two new functions 

\begin{equation}
	\mathcal{N}_{N,k} =2|N|^{k-1/2}\sigma_{1-2k}(N) \, ,
\end{equation}

\noindent and

\begin{equation}
	g_{N,k}(\tau_2) = \sqrt{\tau_2} K_{k-1/2}(2\pi |N|\tau_2) \, .
\end{equation}

\noindent Note that $\mathcal{N}_{-N,k}=\mathcal{N}_{N,k}$ and also $g_{-N,k}(\tau_2)=g_{N,k}(\tau_2)$. We consider the anti-holomorphic derivative first (it will simplify some minus signs)

\begin{align}
	\label{nablaR}
	\overline\nabla \mathcal{E}_k =&\sum_{N\neq0} \mathcal{N}_{N,k} \tau_2^2 (\partial_{\tau_2} - i \partial_{\tau_1}) g_{N,k}(\tau_2) e^{2\pi i N \tau_{1}} \notag \\
	=& \sum_{N\neq0} \mathcal{N}_{N,k} \left((2\pi N \tau_2^2 + \tau_2^2 \partial_{\tau_2}) g_{N,l}(\tau_2) \right) e^{2\pi i N \tau_{1}} \, .
\end{align}

\noindent By defining the operators $\mathcal{O}_1 = 2\pi N \tau_2^2$, $\mathcal{O}_2 = \tau_2^2 \partial_{\tau_2}$ and $\mathcal{O}_+ = \mathcal{O}_1 + \mathcal{O}_2$, we have

\begin{align}
	\overline\nabla^{\gamma} \mathcal{E}_k =&\sum_{N\neq0} \mathcal{N}_{N,k} \left(\mathcal{O}_+^{\gamma} g_{N,k}(\tau_2)\right) e^{2\pi i N \tau_{1}}  \, .
\end{align}

\noindent Similarly we can define $\mathcal{O}_-$ for the holomorphic derivative.

\begin{align}
\nabla^{\gamma} \mathcal{E}_k =&\sum_{N\neq0} \mathcal{N}_{N,k} \left(\mathcal{O}_-^{\gamma} g_{N,k}(\tau_2)\right) e^{2\pi i N \tau_{1}} = \sum_{N\neq0} \mathcal{N}_{N,k} \left(\mathcal{O}_+^{\gamma} g_{N,k}(\tau_2)\right) e^{-2\pi i N \tau_{1}}\, ,
\end{align}

\noindent where $\mathcal{O}_- =- \mathcal{O}_1+\mathcal{O}_2$.  We also used the fact that when we take $N$ to $-N$, $\mathcal{O}_-$ goes to $\mathcal{O}_+$. Based on the binomial expansion given in \cite{wyss2017two} we can write

\begin{align}
	\mathcal{O}_+^{\gamma} = (\mathcal{O}_1+\mathcal{O}_2)^{\gamma} = \sum_{n=0}^{\gamma} \binom{\gamma}{n} \left(\left(\mathcal{O}_1 + {\rm ad}_{\mathcal{O}_2} \right)^{\gamma-n} 1\right) \mathcal{O}_2^{n}\, ,
\end{align}

\noindent where ${\rm ad}_{\mathcal{O}_2} = [\mathcal{O}_2,\bullet]$. For each power of $\tau_2$ we have 

\begin{align}
	{\rm ad}_{\mathcal{O}_2} \tau_2^{n} = n \tau_2^{n+1} \, ,
\end{align}

\noindent in this way, $\left(\left(\mathcal{O}_1 + {\rm ad}_{\mathcal{O}_2} \right)^{\gamma-n} 1\right)$ is a polynomial of $2 \pi N$ and $\tau_2$. The first few examples are

\begin{align}
	&\left(\mathcal{O}_1 + {\rm ad}_{\mathcal{O}_2} \right) 1 = (2\pi N \tau_2)\tau_2 \, ,\notag \\
	&\left(\mathcal{O}_1 + {\rm ad}_{\mathcal{O}_2} \right)^2 1 = (2\pi N \tau_2)^2 \tau_2^2 + 2(2\pi N \tau_2) \tau_2^2 \, , \\
	&\left(\mathcal{O}_1 + {\rm ad}_{\mathcal{O}_2} \right)^3 1 = 6(2\pi N \tau_2) \tau_2^3 + 6(2\pi N \tau_2)^2 \tau_2^3+ (2\pi N \tau_2)^3 \tau_2^3\, . \notag
\end{align}

\noindent As one can see from the examples, we have a recursive relation for the coefficients. By induction, we can prove that this series can be written as

\begin{align}
\left(\mathcal{O}_1 + {\rm ad}_{\mathcal{O}_2} \right)^{\gamma-n} 1 = \tau_2^{\gamma-n} \sum_{a=0}^{\gamma-n} \binom{\gamma-n}{a} \frac{\Gamma{(\gamma-n)}}{\Gamma(a)} (2\pi N \tau_2)^a  \, .
\end{align}

\noindent We also assume that $\frac{\Gamma(0)}{\Gamma(0)} = 1$ so we can extend the above result to $\gamma=n$. Next, we must determine the action of $\mathcal{O}_2^{n} \, g_{N,k}(\tau_2)$.

\begin{align}
	\mathcal{O}_2^n\, g_{N,k}(\tau_2) = \sum_{j=0}^{n} \binom{n}{j} \frac{\Gamma(n+\frac{1}{2})}{\Gamma(j+\frac{1}{2})} \tau_2^{\frac{1}{2}+j+n} K_{k-1/2}^{(j)}(2\pi |N| \tau_2) \, ,
\end{align}

\noindent where $K^{(j)}_{k-1/2}(2\pi |N| \tau_2) = \frac{\dd^{j}}{\dd \tau_2^{j}}K_{k-1/2}(2\pi |N| \tau_2)$ with the following closed formula

\begin{align}
	K^{(j)}_{k-1/2}(2\pi |N| \tau_2) = \left(-\frac{1}{2}\right)^j(2\pi |N| )^{j} \sum_{h=0}^{j}  \binom{j}{h} K_{k-\frac{1}{2}-j+2h}(2\pi |N| \tau_2) \, ,
\end{align}

\noindent Putting all the pieces together, we have

\begin{align}
	\label{NablaCalE}
{\nabla}^{\gamma} \mathcal{E}_k =& \sum_{N\neq0} \mathcal{N}_{N,k} \sum_{n=0}^{\gamma} \sum_{a=0}^{\gamma-n} \sum_{j=0}^{n} \sum_{h=0}^{j} \binom{\gamma}{n}\binom{\gamma-n}{a} \binom{n}{j} \binom{j}{h} \frac{\Gamma(\gamma-n)\Gamma(n+\frac{1}{2})}{\Gamma{(a)}\Gamma(j+\frac{1}{2})}\notag \\
	&\times \left(-\frac{1}{2}\right)^j (2\pi |N| \tau_2)^j (2\pi N \tau_2)^a \tau_2^{\frac{1}{2}+\gamma} K_{k-\frac{1}{2}-j+2h}(2\pi |N| \tau_2) e^{-2\pi i N \tau_1} \, .
\end{align}

\noindent Multiplying this into the anti-holomorphic counterpart and then integrating it gives us the following answer

\begin{align}
	\int_0^{\infty} &\zetas(2m)(m)_{\alpha} \tau_2^{m-\alpha-2\beta-2} R(\tau_2) \, \, \dd \tau_2 =  \\
	& C_{\alpha,\beta}(m,k,l) \zetas(m+k+l-1)\zetas(m-k+l)\zetas(m+k-l)\zetas(-m+k+l) \, , \notag
\end{align}
\noindent where $C_{\alpha,\beta}(m,k,l)$ is 
\begin{align}
	\label{Calphabeta}
C_{\alpha,\beta}(m,k,l) &= \sum_{n_1=0}^{\beta} \sum_{n_2=0}^{\alpha+\beta}\sum_{a_1=0}^{\beta-n_1}\sum_{\substack{a_2=0 \\ a_1+a_2 \in 2\mathbb{N}}}^{\alpha+\beta-n_2} \sum_{j_1=0}^{n_1}\sum_{j_2=0}^{n_2} \sum_{h_1=0}^{j_1}\sum_{h_2=0}^{j_2} \binom{\beta}{n_1}\binom{\beta-n_1}{a_1} \binom{n_1}{j_1} \binom{j_1}{h_1} \notag \\
&\binom{\alpha+\beta}{n_2}\binom{\alpha+\beta-n_2}{a_2} \binom{n_2}{j_2} \binom{j_2}{h_2} (-1)^{j_1 + j_2} 2^{a_1 + a_2} \notag \\
& \times\frac{\Gamma(\beta-n_1)}{\Gamma{(a_1)}}\frac{\Gamma(\alpha+\beta-n_2)}{\Gamma{(a_2)}}\frac{\Gamma(n_1+\frac{1}{2})\Gamma(n_2+\frac{1}{2})}{\Gamma(j_1+\frac{1}{2})\Gamma(j_2+\frac{1}{2})}\frac{\Gamma(m+\alpha)}{\Gamma(m+j_1 + j_2 + a_1 + a_2)}
\notag\\
&\times \left(\frac{m + k - l}{2}\right)_{j_2 + h_1 - h_2 + \frac{a_1 + a_2}{2}} \left(\frac{m - k + l}{2}\right)_{j_1 + h_2 - h_1 + \frac{a_1 + a_2}{2}}\notag \\
&\times  \left(\frac{m + k + l - 1}{2}\right)_{h_2 + h_1 + \frac{a_1 + a_2}{2}} \left(\frac{m - k - l + 1}{2}\right)_{j_1 + j_2 - h_2 - h_1 + \frac{a_1 + a_2}{2}} \, .
\end{align}

\noindent To use this formula, please note that $\frac{\Gamma(\beta-n_1)}{\Gamma{(a_1)}}$ and $\frac{\Gamma(\alpha+\beta-n_2)}{\Gamma{(a_2)}}$ are $\frac{\Gamma(0)}{\Gamma(0)}$ in some terms which should be considered as 1. The first few examples of this function are

\begin{align}
	\begin{split}
	&C_{0,0}(m,k,l) = 1 \, , \\
	&C_{0,1}(m,k,l) = \frac{1}{2}\left({\mu_m-\mu_k-\mu_l}\right) \, , \\
	&C_{1,0}(m,k,l) = \frac{1}{2}\left({\mu_k-\mu_m-\mu_l}\right) \, , \\
	&C_{1,1}(m,k,l) = \frac{1}{2}\left({\mu_l(\mu_m+\mu_k)-(\mu_m-\mu_k)^2}\right) \, ,\\
	&C_{2,0}(m,k,l) = \frac{1}{2}\left({(m-2)_4+(k-2)_4+(l-2)_4-2\mu_k(\mu_m+\mu_l-2)}\right)\, ,\\
	&C_{0,2}(m,k,l) = \frac{1}{2}\left({(m-2)_4+(k-2)_4+(l-2)_4-2\mu_m(\mu_k+\mu_l-2)}\right) \, .
	\end{split}
\end{align}

\noindent Then the full integral is 

\begin{align}
	\label{EEEint}
	\int_{\mathcal{M}_L} &	\tfrac{\nabla^{\alpha} E_m^* \nabla^{\beta} E_k^* \overline\nabla^{\alpha+\beta} E_l^* + {\rm c.c.}}{2\tau_2^{2(\alpha+\beta)}}  \dd\mu =   \sum_{\mathfrak{m},\mathfrak{k},\mathfrak{l}} \zetas(2\mathfrak{m})\zetas(2\mathfrak{k})\zetas(2\mathfrak{l})(\mathfrak{m})_{\alpha}
	 (\mathfrak{k})_{\beta} (\mathfrak{l})_{\alpha+\beta} \tfrac{L^{\mathfrak{m}+\mathfrak{k}+\mathfrak{l}-1}}{\mathfrak{m}+\mathfrak{k}+\mathfrak{l}-1} \notag \\
	&+ C_{\alpha,\beta}(m,k,l) \zetas(m+k+l-1)\zetas(m-k+l) \zetas(m+k-l)\zetas(-m+k+l) \, .
\end{align}
To change \say{=} to \say{$\approx$} we have to take some proper limits since at $m=k+l$, $m=k+l-1$, $k=m+l$, $k=m+l-1$, $l=m+k$, $l=m+k-1$ different terms of (\ref{EEEint}) have simple poles which gets cancelled in the full result. Here we just consider the case where there is no simple pole and the cases $m=k+l$ and $m=k+l-1$. For the first cases without poles, we have

\begin{align}
	&\int_{\mathcal{M}_L} 	\frac{\nabla^{\alpha} E_m^* \nabla^{\beta} E_k^* \overline\nabla^{\alpha+\beta} E_l^* + {\rm c.c.}}{2\tau_2^{2(\alpha+\beta)}}  \dd\mu \approx C_{\alpha,\beta}(m,k,l) \int_{\mathcal{M}_L} \Es_m\Es_k\Es_l \dd\mu \, .
\end{align}
For $m=k+l$ we have

\begin{align}
	\int_{\mathcal{M}_L} &	\frac{\nabla^{\alpha} E_m^* \nabla^{\beta} E_k^* \overline\nabla^{\alpha+\beta} E_l^* + {\rm c.c.}}{2\tau_2^{2(\alpha+\beta)}}\bigg|_{m=k+l}  \dd\mu \approx C_{\alpha,\beta}(m,k,l) \bigg(\int_{\mathcal{M}_L} \Es_m\Es_k\Es_l \dd\mu \notag\\
	&+\zetas(2m-1)\zetas(2k)\zetas(2l) \bigg( \frac{\partial_{m} C_{\alpha,\beta}(m,k,l)}{C_{\alpha,\beta}(m,k,l)}- H_1(m)+H_1(m-\alpha)\bigg)\bigg) \, .
\end{align}
To get this relation, we used 

\begin{align}
	C_{\alpha,\beta}(m,k,l)\bigg|_{m=k+l} = (1-m)_{\alpha} (k)_{\beta} (l)_{\alpha+\beta} \, ,
\end{align}
which is necessary to cancel out the poles and was checked in numerous cases. To differentiate the Pochhammer, we used (\ref{pochref}) and (\ref{pochdif}). Next one is

\begin{align}
	\label{EEEintFinal}
	\int_{\mathcal{M}_L} &	\frac{\nabla^{\alpha} E_m^* \nabla^{\beta} E_k^* \overline\nabla^{\alpha+\beta} E_l^* + {\rm c.c.}}{2\tau_2^{2(\alpha+\beta)}}\bigg|_{m=k+l-1}  \dd\mu \approx C_{\alpha,\beta}(m,k,l) \bigg(\int_{\mathcal{M}_L} \Es_m\Es_k\Es_l \dd\mu \notag\\
	&+\zetas(2m)\zetas(2k-1)\zetas(2l-1) \bigg( -\frac{\partial_{m} C_{\alpha,\beta}(m,k,l)}{C_{\alpha,\beta}(m,k,l)}+ H_1(m+\alpha)-H_1(m)\bigg)\bigg) \, ,
\end{align}
where similarly we used the pole cancellation condition

\begin{align}
	C_{\alpha,\beta}(m,k,l)\bigg|_{m=k+l-1} = (m)_{\alpha} (k-1)_{\beta} (l-1)_{\alpha+\beta} \, .
\end{align}
Other limits have similar results by permuting $m$ ({with} $\alpha$ {inside the harmonic sum}), $k$ ({with} $\beta)$ and $l$ ({with }$\alpha+\beta)$.

{$\blacksquare$} The next function is a shuffle of depth one and an even depth two generalized Eisenstein.

\begin{equation}
	\frac{\nabla^{\alpha} E^*_m\overline\nabla^{\alpha}F^{*+(s)}_{k,l} + {\rm c.c.}}{2\tau_2^{2\alpha}} \, .
\end{equation}

\noindent By unfolding $\nabla^{\alpha} E^*_m$ we have the following zero mode for the seed function

\begin{align}
	f_0(\tau_2) = \, \zetas(2m) (m)_{\alpha} \tau_2^{m+\alpha}\bigg(&\sum_{\mathfrak{k},\mathfrak{l}} \zetas(2\mathfrak{k})\zetas(2\mathfrak{l}) (\mathfrak{k+l})_{\alpha}\frac{\tau_2^{\mathfrak{k}+\mathfrak{l}-\alpha}}{\mu_{\mathfrak{k}+\mathfrak{l}}-\mu_s}
	 + c_{k,l}^{(s)} (1-s)_{\alpha} \tau_2^{1-s-\alpha}\notag \\ &
	+ \frac{1}{\tau_2^{2\alpha}}\mathcal{O}_2^{\alpha}\, R(\tau_2)\bigg) \, ,
\end{align}

\noindent where $R(\tau_2)$ is the rapid decay part of the zero mode of $F_{k,l}^{*+(s)}$, $\mathcal{O}_2$ is defined under equation (\ref{nablaR}) and $c_{k,l}^{(s)}$ is (\ref{hom}). By induction, one can calculate the action of $\mathcal{O}_2^{\alpha}$

\begin{align}
	\label{O2alpha}
	\mathcal{O}_2^{\alpha} \, R(\tau_2) = \sum_{n=0}^{\alpha} \binom{\alpha-1}{n-1} \frac{\alpha!}{n!} \, \tau_2^{\alpha+n} \, \partial_{\tau_2}^n R(\tau_2) \, .
\end{align}
To calculate the contribution of the rapid decay part, we need the following integral

\begin{equation}
	\int_{0}^{\infty} \tau_2^{m+n-2}  \partial_{\tau_2}^{n}R(\tau_2) \dd\tau_2 = (-1)^{n} (m-1)_{n} M(F_{k,l}^{*+(s)};m) \, ,
\end{equation}

\noindent where we have used the integration by part several times and the fact that $R(\tau_2)$ is exponentially suppressed in $\tau_2$. According to (\ref{Feven}), $R(\tau_2)$ satisfies the following differential equation

\begin{equation}
	(\tau_2^2\partial^2_{\tau_2} - \mu_s)R = (E^*_k E^*_l)_R \, ,
\end{equation}

\noindent where $(E^*_k E^*_l)_R$ is the rapid decay part of the zero mode of $E^*_k E^*_l$. It is more convenient to write $R(\tau_2)$ in a recursive form and use the integration by part to find its integral

\begin{equation}
	R(\tau_2) = \frac{1}{\mu_s} \left(\tau_2^2\partial^2_{\tau_2}R(\tau_2) - (E^*_k E^*_l)_R\right) \, ,
\end{equation}

\noindent so for its integral we have

\begin{align}
	M(F_{k,l}^{*+(s)};m)=&  \int_{0}^{\infty} R(\tau_2)\zetas(2m)\tau_2^{m-2} \dd\tau_2\notag\\
	 =& \frac{1}{\mu_s} \bigg(\int_{0}^{\infty} \dd\tau_2 \zetas(2m) \tau_2^{m}\partial_{\tau_2}^2R  - \int_{0}^{\infty} \dd \tau_2 (E^*_k E^*_l)_R \zetas(2m) \tau_2^{m-2} \bigg) \notag \\
	=& \, \frac{1}{\mu_s}\left(\mu_m M (F_{k,l}^{+(s)};m) - \int_{0}^{\infty} \dd \tau_2 (E^*_k E^*_l)_R \zetas(2m) \tau_2^{m-2}\right) \, .
\end{align}

\noindent Now we can write $ M(F_{kl}^{*+(s)};m)$ in the full form for $\mu_m \neq \mu_s$\footnote{For the other cases, we take the limit $m\rightarrow s$ in the final result.}

\begin{align}
	\label{MFplus}
		M (F_{k,l}^{*+(s)};m)&=  \tfrac{1}{\mu_m-\mu_s}\int_{0}^{\infty} \dd \tau_2 (E^*_k E^*_l)_R \zetas(2m) \tau_2^{m-2} = \tfrac{1}{\mu_m-\mu_s} M(E^*_kE^*_l;m)\notag \\
		&= \tfrac{1}{\mu_m-\mu_s}\zetas(m+k+l-1)\zetas(m-k+l)\zetas(m+k-l)\zetas(-m+k+l) \, .
\end{align}

\noindent So we have 

\begin{align}
	\int_0^{\infty}& \zetas(2m) (m)_{\alpha} \tau_2^{m-\alpha-2} \mathcal{O}_2^{\alpha} \, R(\tau_2)  \dd\tau_2 \notag\\
	 =&\,  \frac{1}{\mu_m-\mu_s}\zetas(m+k+l-1)\zetas(m-k+l)\zetas(m+k-l)\zetas(-m+k+l)\notag \\
	&\times \sum_{n=0}^{\alpha} (-1)^{n} (m-1)_n \binom{\alpha-1}{n-1} \frac{\alpha!}{n!}  \\
	= &\, \frac{(1-m)_{\alpha}}{\mu_m-\mu_s}\zetas(m+k+l-1)\zetas(m-k+l)\zetas(m+k-l)\zetas(-m+k+l) \, . \notag
\end{align}
By putting all the pieces together, we have

\begin{align}
	\label{EFplus}
	\int_{\mathcal{M}_L} & \tfrac{\nabla^{\alpha} E^*_m\overline\nabla^{\alpha}F^{*+(s)}_{k,l} + {\rm c.c.}}{2\tau_2^{2\alpha}}  \dd\mu = c_{k,l}^{(s)} (1-s)_{\alpha}\sum_{\mathfrak{m}} \zetas(2\mathfrak{m}) (\mathfrak{m})_{\alpha} \frac{L^{\mathfrak{m}-s}}{\mathfrak{m}-s}  \notag \\
	&+\sum_{\mathfrak{m},\mathfrak{k},\mathfrak{l}} \zetas(2\mathfrak{m})\zetas(2\mathfrak{k})\zetas(2\mathfrak{l})(\mathfrak{m})_{\alpha} (\mathfrak{k+l})_{\alpha}\frac{L^{\mathfrak{m}+\mathfrak{k}+\mathfrak{l}-1}}{(\mathfrak{m}+\mathfrak{k}+\mathfrak{l}-1)(\mu_{\mathfrak{k}+\mathfrak{l}}-\mu_s)} \\
	&+ \tfrac{(m)_{\alpha}(1-m)_{\alpha}}{\mu_m-\mu_s} \zetas(m+k+l-1)\zetas(m-k+l)\zetas(m+k-l)\zetas(-m+k+l) \notag \, .
	\end{align}
There are two sets of limits for this equation where simple poles appear and cancel out. First one is $\mathfrak{m+k+l}=1$ and the other one is $m=s$. These two limits never occur at the same time because it would be out of the spectrum (\ref{Feven}).

\noindent For $m=s$ we have

\begin{align}
	\int_{\mathcal{M}_L} &\tfrac{\nabla^{\alpha} E^*_m\overline\nabla^{\alpha}F^{*+(m)}_{k,l} + {\rm c.c.}}{2\tau_2^{2\alpha}}  \dd\mu \approx \frac{(m)_{\alpha}(1-m)_{\alpha}}{2m-1}\notag \\
	& \times\zetas(m+k+l-1)\zetas(m-k+l)\zetas(m+k-l)\zetas(-m+k+l) \notag\\
	&\times \bigg(\frac{-1}{2m-1}+H_1(m-\alpha)-H_1(m) - \log L -2\frac{{\zetas}'(2m)}{\zetas(2m)}-\frac{{\zetas}'(-m+k+l)}{\zetas(-m+k+l)}\notag\\
	 &\hspace{20pt} +\frac{{\zetas}'(m+k+l-1)}{\zetas(m+k+l-1)}+\frac{{\zetas}'(m-k+l)}{\zetas(m-k+l)}+\frac{{\zetas}'(m+k-l)}{\zetas(m+k-l)}\bigg) \, . 
\end{align}
For $m\neq s$ we have

\begin{align}
	\int_{\mathcal{M}_L} &\tfrac{\nabla^{\alpha} E^*_m\overline\nabla^{\alpha}F^{*+(s)}_{k,l} + {\rm c.c.}}{2\tau_2^{2\alpha}}  \dd\mu \approx \frac{(m)_{\alpha}(1-m)_{\alpha}}{\mu_m-\mu_s}\bigg( \int_{\mathcal{M}_L} \Es_m \Es_k \Es_l \dd\mu \notag \\
	&+ \sum_{\mathfrak{m},\mathfrak{k},\mathfrak{l}} \zetas(2\mathfrak{m})\zetas(2\mathfrak{k})\zetas(2\mathfrak{l}) \bigg(\frac{\mathfrak{m}-1}{\mu_m-\mu_s} + H_1(1-\mathfrak{m}+\alpha)-H_1(1-\mathfrak{m})\bigg) \delta_{\mathfrak{m+k+l},1}\bigg) \, .
\end{align}
By using (\ref{pochref}) and (\ref{pochdif}) we can write the following relation for negative arguments of the harmonic sum

\begin{align}
	H_1(1-{m}+\alpha)-H_1(1-{m}) = H_1(m-\alpha) - H_1(m)
\end{align}

{$\blacksquare$} The next even shuffle function is 

\begin{equation}
	\frac{\nabla^{\alpha}\Es_m\overline\nabla^{\alpha} F_{k,l}^{*-(s)}-{\rm c.c.}}{2\tau_2^{2\alpha}} \, .
\end{equation}

\noindent We follow the steps of the previous integral

\begin{align}
	f_0(\tau_2) =\zetas(2m) (m)_{\alpha} \tau_2^{m-\alpha} \,  \mathcal{O}_2^{\alpha} \, R(\tau_2) \, ,
\end{align}

\noindent where $R(\tau_2)$ is the rapid decay part of the zero mode of $F_{k,l}^{*-(s)}$. We can use (\ref{O2alpha}) to expand this equation. Then we need the following integral

\begin{equation}
	\int_{0}^{\infty} \tau_2^{m+n-2}  \partial_{\tau_2}^{n}R(\tau_2) \dd\tau_2 = (-1)^{n} (m-1)_{n} M(F_{k,l}^{*-(s)};m) \, ,
\end{equation}

\noindent Also, from the definition of $F_{k,l}^{*-(s)}$, $R(\tau_2)$ satisfies the following differential equation

\begin{equation}
	(\tau_2^2\partial^2_{\tau_2} - \mu_s)R(\tau_2) =\left(\frac{\nabla E^*_k \overline\nabla E^*_l -{\rm c.c.}}{2\tau_2^2}\right)_R \, ,
\end{equation}

\noindent where subscript $R$ means that we are just considering the rapid decay part of the zero mode. Once again, we define $R(\tau_2)$ in a recursive form and use the integration by part to integrate

\begin{equation}
	R(\tau_2) = \frac{1}{\mu_s} \left(\tau_2^2\partial^2_{\tau_2}R(\tau_2) - \left(\frac{\nabla E^*_k \overline\nabla E^*_l -{\rm c.c.}}{2\tau_2^2}\right)_R\right) \, ,
\end{equation}

\noindent and then we have

\begin{align}
	\mu_s M(F_{k,l}^{*-(s)};m)= \mu_m M (F_{k,l}^{-(s)};m) 
	- \int_{0}^{\infty} \dd \tau_2 \left(\frac{\nabla E^*_k \overline\nabla E^*_l -{\rm c.c.}}{2\tau_2^2}\right)_R \zetas(2m) \tau_2^{m-2} \, .
\end{align}

\noindent The last integral is zero because from equation (\ref{NablaCalE}) we learn that $\left(\nabla E^*_k \overline\nabla E^*_l\right)_R=\left(\overline\nabla E^*_k \nabla E^*_l\right)_R$. Therefore

\begin{align}
	\label{EFminus}
	\int_{\mathcal{M}_L}	\tfrac{\nabla^{\alpha}\Es_m\overline\nabla^{\alpha} F_{k,l}^{*-(s)}-{\rm c.c.}}{2\tau_2^{2\alpha}} \dd\mu  = 0 \, ,
\end{align}
which means that the integral is exponentially suppressed in $L$.

\subsubsection{Non-shuffle functions}

{$\blacksquare$}  Now we start to integrate the non-shuffle functions. The first even one in table \ref{tableBasis} is $F_{m,k,l}^{*(s)1}$ with the following Laplace equation

\begin{equation}
	(\Delta - \mu_s)F_{m,k,l}^{*(s)1} = E^*_m E^*_k E^*_l \, .
\end{equation}

\noindent We assume that $s\neq 0, 1$. As before, we use the Laplace equation to write the integral as the derivative of the zero mode and the integral of the right-hand side

\begin{align}
	\begin{split}
		\int_{\mathcal{M}_L}F_{m,k,l}^{*(s)1} \dd\mu = \frac{1}{\mu_s} \bigg(\partial_{\tau_2}a_0\bigg|_{\tau_2=L} - \int_{\mathcal{M}_L}{E^*_mE^*_kE^*_l} \dd\mu \bigg) \, .
	\end{split}
\end{align}

\noindent We can obtain the zero mode by solving the following equation

\begin{equation}
	(\tau_2^2\partial^2_{\tau_2} - \mu_s)a_0 = (E^*_mE^*_kE^*_l)_0 \, ,
\end{equation}

\noindent where $(E^*_mE^*_kE^*_l)_0$ is the zero mode of $E^*_mE^*_kE^*_l$. This zero mode has an exponentially suppressed part, which we do not write. The derivative of the rest of the terms is

\begin{align}
	\begin{split}
		\partial_{\tau_2}a_0\bigg|_{\tau_2=L}  = & \sum_{\mathfrak{m},\mathfrak{k},\mathfrak{l}} \zetas(2\mathfrak{m})\zetas(2\mathfrak{k})\zetas(2\mathfrak{l})\frac{(\mathfrak{m}+\mathfrak{k}+\mathfrak{l})L^{\mathfrak{m}+\mathfrak{k}+\mathfrak{l}-1}}{\mu_{\mathfrak{m}+\mathfrak{k}+\mathfrak{l}}-\mu_{s}} \\
		&+ c_{m,k,l}^{(s)1}(s) s L^{s-1} + c_{m,k,l}^{(s)1}(1-s) (1-s)L^{-s}\, ,
	\end{split}
\end{align}

\noindent where the last two terms come from the homogeneous solutions of the Laplace equation. We will not determine the coefficients $c_{m,k,l}^{(s)1}(s)$ and $c_{m,k,l}^{(s)1}(1-s)$ in this work. We can write the last two terms as $\sum_{\mathfrak{s}}{c_{m,k,l}^{(s)1}(\mathfrak{s})\mathfrak{s}L^{\mathfrak{s-1}}}$ where $\mathfrak{s}=\{s,1-s\}$. By adding all the terms, we have

\begin{align}
		\int_{\mathcal{M}_L}F_{m,k,l}^{*(s)1}\dd\mu = & \sum_{\mathfrak{m},\mathfrak{k},\mathfrak{l}} \zetas(2\mathfrak{m})\zetas(2\mathfrak{k})\zetas(2\mathfrak{l})\tfrac{L^{\mathfrak{m}+\mathfrak{k}+\mathfrak{l}-1}}{(\mathfrak{m}+\mathfrak{k}+\mathfrak{l}-1)(\mu_{\mathfrak{m}+\mathfrak{k}+\mathfrak{l}}-\mu_{s})}+\sum_{\mathfrak{s}}\tfrac{c_{m,k,l}^{(s)1}(\mathfrak{s})\mathfrak{s}L^{\mathfrak{s-1}}}{\mu_s}\notag\\&
		 - \frac{1}{\mu_s}\zetas(m+k+l-1)\zetas(m-k+l)\zetas(m+k-l)\zetas(-m+k+l) \notag\\
		\approx & -\frac{1}{\mu_s} 	\left(\int_{\mathcal{M}_L}  \Es_{m}\Es_{k}\Es_{l} \dd\mu +\frac{1}{\mu_s}\sum_{\mathfrak{m},\mathfrak{k},\mathfrak{l}} {\zetas(2\mathfrak{m})\zetas(2\mathfrak{k})\zetas(2\mathfrak{l})}\delta_{\mathfrak{m}+\mathfrak{k}+\mathfrak{l},1} \right)\, .
\end{align}

{$\blacksquare$} The next even function is $F_{m,k,l}^{*(w,s)2+}$ and its Laplace equation is

\begin{equation}
	\label{Fmklws2}
	(\Delta - \mu_w)F_{m,k,l}^{*(w,s)2+} = E^*_m F_{k,l}^{*+(s)} \, ,
\end{equation}

\noindent therefore

\begin{align}
	\begin{split}
		\int_{\mathcal{M}_L}F_{m,k,l}^{*(w,s)2+} \dd\mu = \frac{1}{\mu_w} \bigg(\partial_{\tau_2}a_0\bigg|_{\tau_2=L} - \int_{\mathcal{M}_L}{E^*_mF_{k,l}^{*s+}}\dd\mu\bigg) \, .
	\end{split}
\end{align}

\noindent For the zero mode we have

\begin{align}
	\begin{split}
	(\tau_2^2\partial^2_{\tau_2} - \mu_w)a_0 = (E^*_mF_{k,l}^{*+(s)})_0 = & \sum_{\mathfrak{m},\mathfrak{k},\mathfrak{l}} \zetas(2\mathfrak{m})\zetas(2\mathfrak{k})\zetas(2\mathfrak{l})\frac{\tau_2^{
		\mathfrak{m}+\mathfrak{k}+\mathfrak{l}}}{\mu_{\mathfrak{k}+\mathfrak{l}}-\mu_s}\\
	&+c_{k,l}^{(s)} \sum_{\mathfrak{m}} \zetas(2\mathfrak{m}) \tau_2^{1-s+\mathfrak{m}}\, .
	\end{split}
\end{align}

\noindent Remind that $\mathfrak{m} = \{m, 1-m\}$, $\mathfrak{k} =\{k, 1-k\}$ and $\mathfrak{l} = \{l, 1-l\}$ and $c_{k,l}^{(s)}$ is given in equation (\ref{hom}).

\begin{align}
	\begin{split}
		\partial_{\tau_2}a_0\bigg|_{\tau_2=L}  =& \sum_{\mathfrak{m},\mathfrak{k},\mathfrak{l}} \zetas(2\mathfrak{m})\zetas(2\mathfrak{k})\zetas(2\mathfrak{l})\frac{(\mathfrak{m}+\mathfrak{k}+\mathfrak{l})L^{\mathfrak{m}+\mathfrak{k}+\mathfrak{l}-1}}{(\mu_{\mathfrak{k}+\mathfrak{l}}-\mu_{s})(\mu_{\mathfrak{m}+\mathfrak{k}+\mathfrak{l}}-\mu_w)} \\
		&+  c_{k,l}^{(s)} \sum_{\mathfrak{m}} \zetas(2\mathfrak{m}) \frac{(1-s+\mathfrak{m})L^{\mathfrak{m}-s}}{\mu_{1-s+\mathfrak{m}}-\mu_w} + \sum_{\mathfrak{w}} c_{m,k,l}^{(w,s)2}(\mathfrak{w}) \mathfrak{w} L^{\mathfrak{w}-1}\, ,
	\end{split}
\end{align}

\noindent where the last term is the contribution of the homogeneous solution of (\ref{Fmklws2}), $\mathfrak{w} = \{w,1-w\}$ and $c_{m,k,l}^{(w,s)2}(\mathfrak{w})$'s are the coefficients that we do not determine in this work. From equation (\ref{EFplus}) we know how to integrate $\Es_mF_{k,l}^{*+(s)}$. So the entire integral is

\begin{align}
		\int_{\mathcal{M}_L}F_{m,k,l}^{*(w,s)2+} \dd\mu =& \sum_{\mathfrak{m},\mathfrak{k},\mathfrak{l}}\zetas(2\mathfrak{m}) \zetas(2\mathfrak{k}) \zetas(2\mathfrak{l})\tfrac{ L^{\mathfrak{m}+\mathfrak{k}+\mathfrak{l}-1}}{(\mu_{\mathfrak{k}+\mathfrak{l}}-\mu_s)(\mu_{\mathfrak{m}+\mathfrak{k}+\mathfrak{l}}-\mu_w)(\mathfrak{m}+\mathfrak{k}+\mathfrak{l}-1)} \notag \\
		&+ c_{k,l}^{(s)} \sum_{\mathfrak{m}} \zetas(2\mathfrak{m}) \tfrac{L^{\mathfrak{m}-s}}{(\mathfrak{m}-s)(\mu_{\mathfrak{m}-s}-\mu_w)} + \sum_{\mathfrak{w}} c_{m,k,l}^{(w,s)2}(\mathfrak{w}) \mathfrak{w} \tfrac{L^{\mathfrak{w}-1}}{\mu_w}\\
		&- \tfrac{1}{{\mu_w(\mu_m-\mu_s)}}\zetas(m+k+l-1)\zetas(m-k+l)\zetas(m+k-l)\zetas(-m+k+l)   \, . \notag
\end{align}
Once more, we have two different limits, first one is $m=s$ and the second one is $\mathfrak{m+k+l}=1$ and these two does not occur at the same time because of the spectrum in (\ref{EFplus}).

 For $m=s$ we have
 
\begin{align}
	\int_{\mathcal{M}_L}F_{m,k,l}^{*(w,m)2+} \dd\mu\approx &\, \tfrac{-1}{\mu_w(2m-1)}\zetas(m+k+l-1)\zetas(m-k+l)\zetas(m+k-l)\zetas(-m+k+l) \notag\\
	&\times \bigg(\frac{2m-1-\mu_w}{\mu_w(2m-1)} - \log L -2\frac{{\zetas}'(2m)}{\zetas(2m)}-\frac{{\zetas}'(-m+k+l)}{\zetas(-m+k+l)}\notag\\
	&+\frac{{\zetas}'(m+k+l-1)}{\zetas(m+k+l-1)}+\frac{{\zetas}'(m-k+l)}{\zetas(m-k+l)}+\frac{{\zetas}'(m+k-l)}{\zetas(m+k-l)}\bigg) \, . 
\end{align}

Also, for $m\neq s$ we have

\begin{align}
	\int_{\mathcal{M}_L}F_{m,k,l}^{*(w,s)2+} \dd\mu\approx &\, \tfrac{-1}{\mu_w(\mu_m-\mu_s)}\bigg( \int_{\mathcal{M}_L} \Es_m \Es_k \Es_l \dd\mu -\tfrac{1}{\mu_w(\mu_m-\mu_s)} \sum_{\mathfrak{m},\mathfrak{k},\mathfrak{l}}\zetas(2\mathfrak{m}) \zetas(2\mathfrak{k}) \zetas(2\mathfrak{l}) \notag \\
	&\times \bigg(\mu_m-\mu_s-\mu_w(2\mathfrak{m}-1)\bigg) \delta_{\mathfrak{m+k+l},1} \bigg) \, .
\end{align}

{$\blacksquare$} The next even function is $F_{m,k,l}^{*(s)3+}$ and its Laplace equation is

\begin{equation}
	(\Delta - \mu_s)F_{m,k,l}^{*(s)3+} = \frac{E^*_m\nabla E^*_k\overline{\nabla}E^*_l+ {\rm c.c.}}{2\tau_2^2} \, .
\end{equation}

\noindent Once again, in the case of $\mu_s\neq 0$ we use this Laplace equation to find the integral

\begin{align}
	\begin{split}
		\int_{\mathcal{M}_L}F_{m,k,l}^{*(s)3+} \dd\mu  = \frac{1}{\mu_s} \bigg(\partial_{\tau_2}a_0\bigg|_{\tau_2=L} - \int_{\mathcal{M}_L}\left(\frac{E^*_m\nabla E^*_k\overline{\nabla}E^*_l+ {\rm c.c.}}{2\tau_2^2}\right) \dd\mu \bigg) \, .
	\end{split}
\end{align}

\noindent As we have done before, the first term is

\begin{align}
	\begin{split}
		\partial_{\tau_2}a_0\bigg|_{\tau_2=L}  = \sum_{\mathfrak{m},\mathfrak{k},\mathfrak{l}}  \zetas(2\mathfrak{m})\zetas(2\mathfrak{k})\zetas(2\mathfrak{l})\frac{\mathfrak{k}\mathfrak{l}(\mathfrak{m}+\mathfrak{k}+\mathfrak{l})L^{\mathfrak{m}+\mathfrak{k}+\mathfrak{l}-1}}{\mu_{\mathfrak{m}+\mathfrak{k}+\mathfrak{l}}-\mu_{s}}+ \sum_{\mathfrak{s}} c_{m,k,l}^{(s)3}(\mathfrak{s}) \mathfrak{s} L^{\mathfrak{s}-1} \, ,
	\end{split}
\end{align}

\noindent where the last term is the contribution of the homogeneous solution. Then the integral is

\begin{align}
		\int_{\mathcal{M}_L}&F_{m,k,l}^{*(s)3+} \dd\mu =  \sum_{\mathfrak{m},\mathfrak{k},\mathfrak{l}}\zetas(2\mathfrak{m})\zetas(2\mathfrak{k})\zetas(2\mathfrak{l})\tfrac{\mathfrak{k}\mathfrak{l} \, L^{\mathfrak{m}+\mathfrak{k}+\mathfrak{l}-1}}{(\mu_{\mathfrak{m}+\mathfrak{k}+\mathfrak{l}}-\mu_s)(\mathfrak{m}+\mathfrak{k}+\mathfrak{l}-1)} + \sum_{\mathfrak{s}} c_{m,k,l}^{(s)3}(\mathfrak{s}) \mathfrak{s} \tfrac{L^{\mathfrak{s}-1}}{\mu_s}\notag \\
		&-\tfrac{\mu_m-\mu_k-\mu_l}{2\mu_s}{\zetas}(k+l+m-1){\zetas}(-m+k+l) {\zetas}(m+k-l) {\zetas}(m-k+l) \, .
\end{align}
By taking the limit, we have

\begin{align}
		\int_{\mathcal{M}_L}F_{m,k,l}^{*(s)3+} \dd\mu  \approx  &\frac{-\mu_m+\mu_k+\mu_l}{2\mu_s} \int_{\mathcal{M}_L} \Es_m \Es_k \Es_l \dd\mu \notag \\
		&+	\sum_{\mathfrak{m},\mathfrak{k},\mathfrak{l}}\zetas(2\mathfrak{m})\zetas(2\mathfrak{k})\zetas(2\mathfrak{l}) \bigg(\frac{-\mu_m+\mu_k+\mu_l}{2\mu_s^2}+\frac{2\mathfrak{m}-1}{2\mu_s}\bigg)\delta_{\mathfrak{m+k+l},1} \, .
\end{align}

{$\blacksquare$} The last even function is $F_{m,k,l}^{*(w,s)4+}$ with the following Laplace equation 

\begin{equation}
	(\Delta - \mu_w)F_{m,k,l}^{*(w,s)4+} = \frac{\nabla{E^*_m}{\overline\nabla}F_{k,l}^{*-(s)}-{\rm c.c.}}{2\tau_2^2} \, .
\end{equation}

\noindent The integral can be written as

\begin{align}
	\begin{split}
		\int_{\mathcal{M}_L}F_{m,k,l}^{*(w,s)4+} \dd\mu = \frac{1}{\mu_w} \bigg(\partial_{\tau_2}a_0\bigg|_{\tau_2=L} - \int_{\mathcal{M}_L}\frac{\nabla{E^*_m}{\overline\nabla}F_{kl}^{*s-}- {\rm c.c.}}{2\tau_2^2}\dd\mu\bigg).
	\end{split}
\end{align}

\noindent Since $F_{kl}^{*-(s)}$ does not have a Laurent polynomial, $\partial_{\tau_2}a_0\bigg|_{\tau_2=L}$ just has the contribution of the homogeneous term. In addition, equation (\ref{EFminus}) tells us that the second term is also exponentially suppressed. So we have

\begin{align}
	\begin{split}
		\int_{\mathcal{M}_L}F_{m,k,l}^{*(w,s)4+} \dd\mu = \sum_{\mathfrak{s}} c_{m,k,l}^{*(w,s)4+}(\mathfrak{s}) \, \mathfrak{s}\, L^{\mathfrak{s}-1} \,  \approx 0 \, .
	\end{split}
\end{align}

\section{Conclusion}

In this paper, we used the depth filtration of modular graph forms and evaluated the integral of their basis up to depth three over $\mathcal{M}_L$. We also proved three theorems that can be useful in finding the direct integral of the functions over $\mathcal{M}_L$ without regularizing them with the non-holomorphic Eisenstein series. We can use methods like the sieve algorithm to expand any MGF in terms of $\beta^{eqv}$s and, finally, in terms of the basis presented in this paper. This method can also be applied to the generating function of the closed string amplitude \cite{Gerken:2020yii}; it is versatile enough for any number of punctures, not just the MGFs that appear in the four-graviton Type II superstring theory in ten-dimensional Minkowski spacetime. Additionally, the space of the functions spanned by this basis is larger than the space of MGFs \cite{Dorigoni:2021jfr, Dorigoni:2021ngn}.

The apparent next step is to integrate a depth four and higher depth basis. However, even if we already had a higher depth basis with a suitable Laplace system, which we do not currently have, as demonstrated in this paper, we must first find the integral of the product of four and more non-holomorphic Eisenstein series, a task yet to be accomplished. 

The lowest transcendental weight we can have at depth $d$ is $2d$. It implies that we should be able to express any MGF of weight $w$ in terms of a basis up to weight $w/2$. Therefore, all the MGFs of weight seven can be represented in terms of the basis that we integrated in this paper (plus the one with eigenvalue zero, which will be integrated in the future work \cite{Claasen2024appear}). It means that the term $D^{14}\mathcal{R}^4$ arising in the low-energy expansion of four gravitons can potentially be found through this paper, a topic for future research \cite{Claasen2024appear}.

Another path to pursue is to return to the generating function of \cite{Gerken:2020xte} and integrate it over $\mathcal{M}_L$. Since this generating function was written for any number of punctures, it offers an opportunity to study the integral of all the one-loop closed (super)string amplitudes over $\mathcal{M}_L$. The investigation of the integral of the generating function will also be a subject of future work \cite{Doroudiani2023appear}.

\textbf{Acknowledgment:}
I am grateful to Axel Kleinschmidt for the extensive discussions and insights that greatly influenced this work and for many fruitful comments on the draft. I also thank Oliver Schlotterer, Daniele Dorigoni, Guillaume Bossard, Federico Zerbini and Emiel Claasen for valuable discussions. Special thanks to Oliver Schlotterer and Federico Zerbini for their comments on the first version of the paper and I thank the anonymous referee for valuable comments and suggestions. My appreciation goes to École Polytechnique for its hospitality during the early stages of this research. This work was supported by the IMPRS for Mathematical and Physical Aspects of Gravitation, Cosmology, and Quantum Field Theory.

\end{document}